\documentclass[11pt,a4paper]{article}

\usepackage[dvips]{graphicx}
\usepackage{amsfonts,amssymb,eucal,amsmath,epsfig}

\newtheorem{theorem}{Theorem}[section]
\newtheorem{example}[theorem]{Example}
\newtheorem{lemma}[theorem]{Lemma}
\newtheorem{proposition}[theorem]{Proposition}
\newtheorem{definition}{Definition}[section]
\newtheorem{remark}{Remark}[section]

\begin{document}

\begin{center}
\textbf{New method for the Numerical Calculation of Hydrodynamics
Shocks}

\emph{Dedicated to Tatsiana  Radyna}
\end{center}

\begin{center}
\textbf{Mikalai Radyna}\footnote{e-mail:~ mik\_ radyna@yahoo.com}
\end{center}

\begin{center}
Institute of Mathematics,

National Academy of Sciences of Belarus,

Surganova 11,  Minsk, 220072, Belarus

e-mail: kolya@im.bas-net.by
\end{center}

\begin{abstract}
The equations of  hydrodynamics are rewritten  in sense of
functionals with values in Non-Archimedean field of Laurent series
or  $\mathbf{R}\langle\varepsilon\rangle$-distributions. A new
ideology for understanding of conservation laws is proposed. A set
of nonlinear algebraic equations suitable for the numerical work
is given. The Newton iteration method are used for calculation of
microscopic shock profiles for the equations of compressible flow.
\end{abstract}

\noindent KEY WORDS: generalized functions, distributions,
algebra, Hermite functions, conservation law, Hopf equation,
equations of compressible flow, soliton, shock wave. \medskip

 \noindent PACS numbers 02.30.Sa; 02.30.Mv; 02.60.-x; 52.35.Tc


\section{Historical remark and Introduction}

In 1943-44, von Neumann became convinced that the calculation of
the flows of compressible fluids containing strong shocks could be
accomplished only by numerical methods. He conceived the idea of
capturing shocks, i.e., of ignoring the presence of a
discontinuity. Employing a Lagrangian description of compressible
flow, setting heat conduction and viscosity equal to zero, von
Neumann replaced space and time derivatives by symmetric
difference quotients. Calculations using this scheme were carried
out; the approximation resulting from these calculations (see
\cite{Neumann}) showed oscillations on the mesh scale behind the
shock. Von Neumann boldly conjectured that the oscillations in
velocity represent the heat energy created by the irreversible
action of the shock, and that as $\Delta x$ and $\Delta t$ tend to
zero, the approximate solutions tend in the weak sense to the
discontinuous solution of the equations of compressible flow.

In \cite{Lax0} it was counterconjectured  that von Neumann was
wrong in his surmise, i.e., that although the approximate
solutions constructed by his method do converge weakly, the weak
limit fails to satisfy the law of conservation of energy.

In \cite{Goodman-Lax} J.Goodman and P.Lax investigated von
Neumann's algorithm applied to the scalar equation
\begin{equation}\label{HopfEquation}
u_t+uu_x=0
\end{equation}
(it is called the Hopf equation \cite{Hopf}), in the semidiscrete
case. Using numerical experimentation and analytical techniques
the demonstrated the weak convergence of the oscillatory
approximations, and that the weak limit fails to satisfy the
scalar equation in question.

Von Neumann's dream of capturing shocks was realized in his joint
work with Richtmyer in 1950, see \cite{Neumann-Richtmyer}.
Oscillations were eliminated by the judicious use of artificial
viscosity; solutions constructed by this method converge uniformly
except in a neighborhood of shocks, where they remain bounded and
are spread out over a few mesh intervals. The limits appear to
satisfy the conservation laws of compressible flow. The
conservation of mass and momentum is the consequence of having
approximated these equations by difference equations in
conservation form; but the von Neumann-Richtmyer difference
approximation to the energy equation is not in conservation form.

In the paper  \cite{Hou-Lax} T.Hou and P.Lax compared the results
of a von Neumann-Richtmyer calculation with the weak limit of of
calculations performed by  von Neumann's original method.

The aim of this paper, at first, it is to introduce the idea of
understanding of conservation laws; the second is to proposed a
numerical method for calculation of hydrodynamic shocks profile
\emph{without using differences schemes}.

\begin{figure}[ht]
\centering\includegraphics[width=0.5\linewidth]{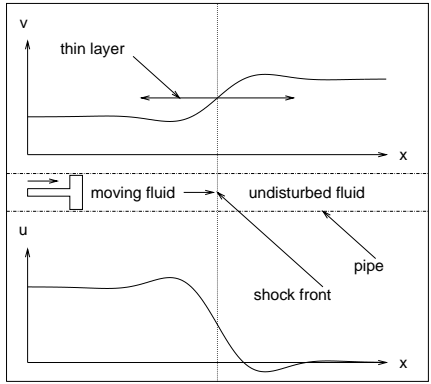}
\caption{\footnotesize Steady-state plane shock}\label{fi0}
\end{figure}

We believe that there is no discontinues in the nature of shocks.
Discontinues solution of hydrodynamic equations only a rough
mathematical model of shocks. When viscosity is taken into
account, for example, the shocks are seen to be smeared out, so
that the pure mathematical surface of discontinuety are replaced
by the thin layers ($10^{-7}$--$10^{-8}$ m) in which pressure,
density, temperature and etc. vary rapidly but continuosly (see
Fig.\ref{fi0}).

The equations of compressible flow in one space dimension can be
written in the following Lagrangian form:
\begin{equation}\label{system0}
\begin{array}{l}
u_t+p_x=0 \\ v_t-u_x=0 \\ e_t+pu_x=0
\end{array}
\end{equation}
It is a classical approach. Here $u$ is velocity, $p$ is pressure,
$v$ specific volume, and $e$ internal energy, connected with $p$
and $v$ via an equation of state. Here we mostly study the case
when the equation of state is given by the $\gamma$-law with
$\gamma=1.4$ $$e=\frac{pv}{\gamma -1}.$$ Substituting that into
the (\ref{system0}) we get the following equation
\begin{equation}\label{work'}
\frac{1}{\gamma -1} \left(pv \right)_t+pu_x=0.
\end{equation}
The first equation  of (\ref{system0}) is conservation of
momentum, the second equation conservation of mass, the third the
work equation.

We will consider a steady-state shock. Imagine a long pipe
containing a fluid initially in equilibrium (thermally and
mechanically), into which a piston is pushing from one end with
constant speed, as shown in Fig.\ref{fi0}. In the presence of
dissipation the specific volume, $v$, and the fluid  velocity, $u$
are as shown by the curves. \emph{Our problem is to calculate the
exact shape of this curves.} The shock is steady, at least
approximately, after it has gone to a sufficiently great distance
from the initiating piston. Then $u, v, e$, etc. depend on $x$ and
$t$ only through the combination $y=x-ct,$ where $c$ is a speed of
the shock relative to the original, or Lagrangian, coordinates.

Now, we are going to study the equations (\ref{system0}) in
specific sense. Namely, we are going to rewrite mentioned
equations in the sense of
$\mathbf{R}\langle\varepsilon\rangle$--distributions. We give the
definition of the special kind of solutions of the some
conservation laws in the sense of
$\mathbf{R}\langle\varepsilon\rangle$-distributions and consider
the method for the numerical calculations of the smooth shocks and
soliton like solutions of the  Hopf equation and equations of
compressible flow in the mentioned sense. This method based on
orthogonal system of the Hermite functions as a base for
calculation of such solutions (i.e. shocks and infinitely narrow
solitons). Calculations of profiles of infinitely narrow soliton
and shock wave are reduced to the nonlinear system of algebraic
equations in $\mathbf{R}^{n+1}$, $n>1$. We proved, using the
Schauder fixed point theorem \cite{Schauder}, that the mentioned
system has at least one solution in $\mathbf{R}^{n+1}$. We showed
that there is possibility to find out some of the solutions of
this system using the Newton iteration method
\cite{Kantorovich-Akilov}. We considered examples and numerical
tests. We also should emphasis that proposed numerical approach do
not use a difference scheme (see also \cite{Radyna}).

Let us consider a bit of theory which we will apply to
conservation laws.

\section{Non-Archimedean field of Laurent series and
$\mathbf{R}\langle\varepsilon\rangle$--distributions.} The theory
of Non-Archimedean fields was considered in the book by
A.H.Light\-stone and A.Robin\-son  \cite{Lightstone-Robinson}.

\begin{definition}\label{definition1}
A \emph{Laurent series} is a formal object
$$\sum_{n=0}^{\infty}
\xi_{n+k} \varepsilon^{n+k}$$ where $k$ is a fixed (i.e., fixed
for this Laurent series), each $\xi_i\in\mathbf{R}$, and either
$\xi_k\not= 0$ or each $\xi_i=0$.
\end{definition}

The Laurent series $\sum_{n=0}^{\infty} \xi_n \varepsilon^n $,
where $\xi_0=1$ and $\xi_n=0$ if $n>0$, is denoted by 1. It is
easy to see that the Laurent series is a field. Let us denote it
by $\mathbf{R}\langle\varepsilon\rangle$.  The norm on the field
of Laurent series can define $$|x|_\nu=e^{-\nu (x)}\quad
\textrm{for}\quad \textrm{each}\quad x\in
\mathbf{R}\langle\varepsilon\rangle $$ (in place of $e$ can use
any number greater than 1). The function $\nu(x)$ is a
Non-Archimedean valuation. Define $$ \nu(0)=\infty
\quad\textrm{and}\quad \nu\left(\sum_{n=0}^{\infty} \xi_{n+k}
\varepsilon^{n+k}\right)=k \quad\textrm{if}\quad
\sum_{n=0}^{\infty} \xi_{n+k} \varepsilon^{n+k}\not=0,\quad
\xi_k\not= 0.$$

The norm $|\cdot|_\nu$ have properties
\begin{enumerate}
\item $|x|_\nu=0 \quad\textrm{if and only if}\quad x=0,$

\item $|xy|_\nu=|x|_\nu\cdot |y|_\nu,$

\item $|x+y|_\nu\leq\max\left\{|x|_\nu,\, |y|_\nu\right\}.$
\end{enumerate}
Here, we propose a general construction of the
$\mathbf{R}\langle\varepsilon\rangle$--valued generalized
functions \cite{Radyno}, \cite{Radyno1}. These objects are a
natural generalization of Sobolev-Schwartz distributions. We call
them as $\mathbf{R}\langle\varepsilon\rangle$--distributions.

\begin{enumerate}
\item Consider all functions $f(x,\varepsilon)\in
C^{\infty} (\mathbf{R}\times (0,1))$ such that integrals
$$\int\limits_{-\infty}^{+\infty} f(x,\varepsilon )\psi (x)dx$$
exist for any  $\varepsilon$ and for all $\psi (x)$  from a given
class of functions $\mathcal{X}$ ($\mathcal{X}$ can be $C^\infty_0
(\mathbf{R}),$ $\mathcal{S}(\mathbf{R})$ and etc.).

\item   Suppose also  that $\displaystyle\int\limits_{-\infty}^{+\infty}
f(x,\varepsilon )\psi (x)dx$ is a number $a_{f,\varepsilon}(\psi)$
from the field of Laurent series
$\mathbf{R}\langle\varepsilon\rangle$.

\item The two functions  $f(x,\varepsilon )$ and $g(x,\varepsilon )$
call equivalent with respect to test functions $\mathcal{X}$ if
and only if $$\int\limits_{-\infty}^{+\infty} f(x,\varepsilon
)\psi
(x)dx=a_{f,\varepsilon}(\psi)=a_{g,\varepsilon}(\psi)=\int\limits_{-\infty}^{+\infty}
g(x,\varepsilon )\psi (x)dx.$$ The equality means in sense of the
field of Laurent series $\mathbf{R}\langle\varepsilon\rangle$ for
all functions $\psi\in\mathcal{X}$. Classes of equivalent
functions call
$\mathbf{R}\langle\varepsilon\rangle$\emph{--functions}. The
expression $$\int\limits_{-\infty}^{+\infty} f(x,\varepsilon )\psi
(x)dx$$ associates a number from
$\mathbf{R}\langle\varepsilon\rangle$ with every $\psi$. Such a
quantity is called a functional. In this case a linear functional
map $\mathcal{X}$ into the Non-Archimedean field
$\mathbf{R}\langle\varepsilon\rangle$. Call these functionals as
$\mathbf{R}\langle\varepsilon\rangle$\emph{-distributions}.
\end{enumerate}
Thus,
\begin{proposition}
$\mathbf{R}\langle\varepsilon\rangle$--function
$f(x,\varepsilon)=0$ if and only if
$$\int\limits_{-\infty}^{+\infty} f(x,\varepsilon )\psi (x)dx=0\in
\mathbf{R}\langle\varepsilon\rangle$$ for every $\psi$ from
$\mathcal{X}$.

The set of all
$\mathbf{R}\langle\varepsilon\rangle$\emph{--distributions} denote
by $\mathcal{R}(\mathcal{X})$
\end{proposition}
\begin{remark}
Recall that the idea of representation of a function $f\in L_{\it
loc }^1(\mathbf{R})$ in terms of a linear functional
$$C_0^{\infty}(\mathbf{R})\ni\psi\longmapsto
\int\limits_{-\infty}^{+\infty} f(x)\psi (x)dx\in\mathbf{R} $$
based on well-known proposition that if  $f\in L_{\it loc
}^1(\mathbf{R})$ and $\displaystyle\int\limits_{-\infty}^{+\infty}
f(x)\psi (x)dx=0$ for any $\psi\in C_0^{\infty}(\mathbf{R})$ then
$f=0$ almost everywhere.
\end{remark}

Let us consider an example of the
$\mathbf{R}\langle\varepsilon\rangle$-distribution.

\begin{example}
Take $\mathcal{X}=C_0^{\infty}(\mathbf{R})$ and $f(x,\varepsilon
)=\varphi (x/\varepsilon)$,  $\varphi (x)\in
C_0^{\infty}(\mathbf{R})$ then
$\mathbf{R}\langle\varepsilon\rangle$--distribution can write in
the following form. $$\int\limits_{-\infty}^{+\infty} \varphi
(x/\varepsilon)\psi (x)dx=\varepsilon
\int\limits_{-\infty}^{+\infty} \varphi (x) dx \psi
(0)+\varepsilon^2 \displaystyle\int\limits_{-\infty}^{+\infty}
x\varphi (x) dx \displaystyle\frac{\psi' (0)}{1!}+\ldots .$$

Note that  $\varphi (x/\varepsilon)$ converges to the function
$$u(x)=\left\{
\begin{array}{ll}
\varphi (0), & \textrm{if} \,\, x=0,\\ 0, & \textrm{if}\,\,
x\not=0.\\
\end {array} \right.$$ Last function almost everywhere equals to zero.

Like Sobolev-Schwartz distributions we can \emph{differentiate}
$\mathbf{R}\langle\varepsilon\rangle$--distributions. For example,
$$\displaystyle\int\limits_{-\infty}^{+\infty} \frac{d}{dx}\varphi
(x/\varepsilon)\psi
(x)dx=-\displaystyle\int\limits_{-\infty}^{+\infty} \varphi
(x/\varepsilon)\frac{d}{dx}\psi (x)dx,$$

$$-\displaystyle\int\limits_{-\infty}^{+\infty} \varphi
(x/\varepsilon)\frac{d}{dx}\psi (x)dx=-\varepsilon
\displaystyle\int\limits_{-\infty}^{+\infty} \varphi (x) dx \psi'
(0)-\varepsilon^2 \displaystyle\int\limits_{-\infty}^{+\infty}
x\varphi (x) dx \displaystyle\frac{\psi'' (0)}{1!}-\ldots .$$
\end{example}

It  is evident that
$\mathbf{R}\langle\varepsilon\rangle$--distributions are more
general objects than Sobolev-Schwartz distributions
\cite{Schwartz}, \cite{Sobolev}.

\section{Conservation laws. Non-Archimedean approach.} A
conservation law asserts that the rate of change of the total
amount of substance contained in a fixed domain $G$ is equal to
the flux of that substance across the boundary of $G$. Denoting
the density of that substance by $u$, and the flux by $f$, the
conservation law is $$\frac{d}{dt}\int_{G} u(t,x) d x = -
\int_{\partial G} f \cdot \vec{n} dS.$$ Applying the divergence
theorem and taking $d/dt$ under the integral sign we obtain
$$\int_G (u_t+\mathbf{div} f) dx=0.$$ Dividing by vol $(G)$ and
shrinking $G$ to a point where all partial derivatives of $u$ and
$f$ are continuous we obtain the differential conservation law
$$u_t(t,x)+\mathbf{div} f (u(t,x))=0.$$ Note, that if $f(u)=u^2/2$
then we obtained the Hopf equation (\ref{HopfEquation}). In
general, previous calculations based on the following well known
proposition.

\begin{proposition}
If  $G\in L_{\it loc }^1(\mathbf{R})$ and
$\displaystyle\int\limits_{-\infty}^{+\infty} G(x)\psi (x)dx=0$
for any $\psi\in C_0^{\infty}(\mathbf{R})$ then $G=0$ almost
everywhere.
\end{proposition}

\begin{definition}\label{definition2}
Let us consider two sets of the smooth functions, depending on a
small parameter $\varepsilon \in (0,1]$. Let us take all functions
$v(t,x,\varepsilon)$ which have the type
$$v(t,x,\varepsilon)=l_0+\Delta l \varphi
\left(\frac{x-ct}{\varepsilon}\right),$$ $l_0, \Delta l, c$ are
real numbers, $\Delta l\not=0$ and $\varphi\in
\mathcal{S}(\mathbf{R} )$,
$\displaystyle\int\limits_{-\infty}^{+\infty} \varphi(y)dy=1$. We
denote this set of functions by $I$. We call $I$ as a set of
infinetely narrow solitons.
\end{definition}

\begin{definition}\label{definition3}
Now, let us take all functions $w(t,x,\varepsilon)$ which have the
type $$w(t,x,\varepsilon)=h_0+\Delta h H
\left(\frac{x-at}{\varepsilon}\right),$$ $h_0, \Delta h, a$ are
real numbers, $\Delta h\not=0$ and
$H(x)=\displaystyle\int\limits_{-\infty}^{x} \theta (y) d y$,
$\displaystyle\int\limits_{-\infty}^{+\infty} \theta (y) d y=1$
and $\theta\in \mathcal{S}(\mathbf{R} )$. We denote this set of
functions by $J$. We call $J$ as a set of shock waves.
\end{definition}

It is natural to consider conservation laws as an integral
expressions which contain the time $t$ as parameter. Therefore, we
introduce the following concept.

\begin{definition}\label{definition4}
The function $v\in I$ (or $w\in J$) will be a solution of the Hopf
equation up to $e^{-l}$, $l\in \mathbf{N}_0$ in the sense of
$\mathbf{R}\langle\varepsilon\rangle$--distributions if for any
$t\in [0,T]$

\begin{equation}\label{solution-1}
\int\limits_{-\infty}^{+\infty} \left\{v_t(t,x,\varepsilon )+
v(t,x,\varepsilon )v_x(t,x,\varepsilon ) \right\}\psi (x)
dx=\displaystyle\sum\limits_{k=l}^{+\infty} \xi_k \varepsilon^k\in
\mathbf{R}\langle\varepsilon\rangle,
\end{equation}

\begin{equation}\label{solution-2}
\int\limits_{-\infty}^{+\infty}\left\{w_t(t,x,\varepsilon )+
w(t,x,\varepsilon )w_x(t,x,\varepsilon ) \right\}\psi (x)
dx=\displaystyle\sum\limits_{k=l}^{+\infty} \eta_k
\varepsilon^k\in \mathbf{R}\langle\varepsilon\rangle
\end{equation}
for every $\psi\in\mathcal{S}(\mathbf{R}).$ In case when $l$ is
equal to  $+\infty$  the function $v (t,x,\varepsilon )$ (or $w
(t,x,\varepsilon )$) exactly satisfies the Hopf equation in the
sense of $\mathbf{R}\langle\varepsilon\rangle$--distributions.
\end{definition}

Certainly, one can consider instead of the Hopf equation some
conservation law.

From mathematical point of view, we deal with a infinitely
differentiable functions in definitions \ref{definition2} and
\ref{definition3}, so that we avoid the problem of distribution
multiplication. From physical point of view, functions from the
set $I$ or $J$ can describe fast processes. Mathematical models of
such processes based on functions from $I$ or $J$ may give
additional information and take in account a short zone where
physical system make a jump from one position to another.

Thus, we will consider solutions of the Hopf equation which are
infinitely narrow solitons or shock waves. It easy to see that

$$v(t,x,\varepsilon )\longrightarrow \left\{\begin{array}{ll}
l_0+\Delta l \varphi (0),& if \quad x=ct , \\ l_0 , & if \quad
x\not=ct .\end{array}\right. \quad \textrm{as} \quad
\varepsilon\to 0$$

$$w(t,x,\varepsilon )\longrightarrow h_0 +\Delta h H (x-at), \quad
\textrm{as} \quad  \varepsilon\to 0$$ $H$ is Heaviside function.

\section{Method for the numerical calculations of the
microscopic profiles of soliton like solutions of the Hopf
equation in the sense of
$\mathbf{R}\langle\varepsilon\rangle$--distributions.}

Thus, conservation laws are integral expressions. Therefore, it is
natural, that one can interpret the Hopf equation in the sense of
the definition \ref{definition4}.

We will seek a solution of the Hopf equation in the type of
infinitely narrow soliton, i.e. let us $v\in I$. Substitute
$v(t,x,\varepsilon)$ into integral expression (\ref{solution-1})
using  the following formulas
\begin{equation}
\int\limits_{-\infty}^{+\infty}\frac{\partial}{\partial
t}\left\{\varphi \left(\frac{x-c
t}{\varepsilon}\right)\right\}\psi (x)
dx=\sum\limits_{k=0}^{+\infty} c\varepsilon^{k} m_k
\frac{1}{k!}\psi^{(k+1)}(ct),
\end{equation}
\begin{equation}
\int\limits_{-\infty}^{+\infty}\frac{\partial}{\partial x
}\left\{\frac{1}{2}\varphi^2\left(\frac{x-c
t}{\varepsilon}\right)\right\} \psi (x) d
x=\sum\limits_{k=0}^{+\infty} -\varepsilon^{k} g_k
\frac{1}{k!}\psi^{(k+1)}(ct).
\end{equation}
We denote
\begin{equation}
m_k(\varphi)=\int\limits_{-\infty}^{+\infty} y^k\varphi(y) d y,
\,\,g_k(\varphi)=\int\limits_{-\infty}^{+\infty}
y^k\frac{\varphi^2(y)}{2} d y, \,\, k=0,1,2,\ldots .
\end{equation}
Thus, we obtain
\begin{equation}\label{Laurent series}
\int\limits_{-\infty}^{+\infty}\left\{ v_t +v v_x \right\}\psi  d
x=\sum\limits_{k=0}^{+\infty} \left\{\Delta l(c-l_0) m_k -(\Delta
l)^2 g_k\right\}\varepsilon^{k} \frac{\psi^{(k+1)}(c t)}{k!}.
\end{equation}
From the last expression we have conditions for the function
$\varphi(x)$. Namely,
\begin{equation}\label{pre-conditions}
 g_k(\varphi) - \frac{c-l_0}{\Delta l} m_k(\varphi)=0,
\,\, k=0,1,2 \ldots.
\end{equation}
From the first  ($k=0$) we have
\begin{equation}
\frac{c-l_0}{\Delta
l}=\frac{g_0}{m_0}=\frac{1}{2}\int\limits_{-\infty}^{+\infty}
\varphi^2(x) d x.
\end{equation}
Hence, we can rewrite conditions (\ref{pre-conditions}) as
follows.
\begin{equation}\label{pre-conditions-1}
\int\limits_{-\infty}^{+\infty} \varphi^2(x)d x \cdot
\int\limits_{-\infty}^{+\infty} x^k \varphi(x)d
x=\int\limits_{-\infty}^{+\infty} x^k \varphi^2(x)d x, \,\,k=0,1,2
\ldots.
\end{equation}
Now, let us prove the following lemma.
\begin{lemma}
For any non-negative integer $n$  exists such function
$\varphi\in\mathcal{S}(\mathbf{R})$, $\varphi\not\equiv 0$ which
satisfies the following system of non-linear equations:
\begin{equation}\label{conditions}
\int\limits_{-\infty}^{+\infty} x^k \varphi(x)d
x=\int\limits_{-\infty}^{+\infty} x^k \varphi^2(x)d x /
\int\limits_{-\infty}^{+\infty} \varphi^2(x)d x \,\,\,\,k=0,1,2
\ldots n.
\end{equation}
\end{lemma}
{\it Proof.} First, we will seek function  $\varphi(x)$ in the
following type:
\begin{equation}\label{function}
\varphi (x)=c_0 h_0(x)+c_1 h_1(x)+\ldots+c_{n} h_{n}(x),
\end{equation} where
\begin{equation}
h_{k}(x)=\frac{H_{k}(x)}{\sqrt{2^k k!}\sqrt[4]{\pi}} e^{-x^2/2}
\,\,\mathrm{are} \,\, \mathrm{Hermit} \,\, \mathrm{functions.}
\end{equation}
Then we substitute the expression (\ref{function}) into conditions
(\ref{conditions}). After that we will have nonlinear system of
$n+1$ equations with $n+1$ unknowns ($c_0, c_1, c_2, \ldots ,
c_n$). We write this system by the following way.

\begin{equation}\label{system}
A\vec{x}=\mathcal{N}(\vec{x}), \,\,\, \vec{x}=(c_0, c_1, \ldots ,
c_n)
\end{equation}
$A$ is a matrix with elements
$$A_{kj}=\int\limits_{-\infty}^{+\infty} x^k h_j (x) d x =
(-\mathbf{i})^j \mathbf{i}^k \sqrt{2\pi} h_j^{(k)}(0),
\,\,\,\mathbf{i}=\sqrt{-1},\,\,\, k,j=0,1,2,\ldots n $$

$\mathcal{N}$ is nonlinear map such that
\begin{equation}
\mathcal{N}(\vec{x})=\frac{1}{\|\vec{x}\|^2}\sum\limits_{k=0}^n
(N(k)\vec{x},\vec{x})\vec{e}_k\equiv \sum\limits_{k=0}^n
f_k(\vec{x})\vec{e}_k
\end{equation}
Vector $\vec{e}_k=(e_0, e_1, \ldots , e_n)$ such that $e_k=1$ and
$e_j=0$ for all $j\neq k$. $N(k)$ are matrices with elements
\begin{equation}
N_{ij}(k)=\int\limits_{-\infty}^{+\infty} x^k h_i(x) h_j(x) d x
,\,\,\, i,j,k=0,1,2 \ldots n
\end{equation}
and functions $$f_k(\vec{x})=\frac{(N(k)
\vec{x},\vec{x})}{\|\vec{x}\|^2}.$$ Note that functions
$f_k(\vec{x})$ are continuous everywhere except $\vec{x}=0$ and
$|f_k(\vec{x})|\leq \|N(k)\|$ due to Cauchy-Bunyakovskii
unequality. Matrix $A$ is invertible for any $n$ because of $\det
(A)$ is a Wronskian for the linear independent system of Hermit
functions $h_0(x)$, $h_1(x)$, ... $h_n(x)$ and $$\det
(A)=(2\pi)^\frac{(n+1)}{2} W(h_0(0), h_1(0), \ldots h_n(0)).$$ We
can write the system (\ref{system}) as
\begin{equation}\label{functionF}
\vec{x}=\sum\limits_{k=0}^n f_k(\vec{x})A^{-1}\vec{e}_k\equiv
F(\vec{x}) \,\,\,\mathrm{or}\,\,\, \vec{x}=A^{-1}
(\mathcal{N}(\vec{x}))\equiv F(\vec{x}).
\end{equation}
Let us describe the function $F:\mathbf{R}^{n+1} \longmapsto
\mathbf{R}^{n+1}$. It is continuous except $\vec{x}=0$ and
bounded. Indeed,
\begin{equation}
\|F(\vec{x})\|\leq \|A^{-1}\| \sum\limits_{k=0}^n \|N(k)\|, \,\,\,
r_n=\|A^{-1}\| \sum\limits_{k=0}^n \|N(k)\|.
\end{equation}
Let us consider function $\mathcal{N}(\vec{x})$. It is continuous
function everywhere in $\mathbf{R}^{n+1}$ except $\vec{x}=0$ and,
moreover, $\mathcal{N}(\mathbf{R}^{n+1}\backslash
\{0\})\subset\Pi_1$ where $\Pi_1=\{\vec{z}\in\mathbf{R}^{n+1}:
z_0=1 \}$ is a plane. Further $A^{-1}(\Pi_1)=\Pi_2$ where
$\Pi_2=\{\vec{y}\in\mathbf{R}^{n+1}: \sum\limits_{k=0}^n a_{0j}
y_j=1 \}$ is another plane.
$$A_{0j}=\int\limits_{-\infty}^{+\infty} h_j (x) d x =
(-\mathbf{i})^j \sqrt{2\pi} h_j(0),
\,\,\,\mathbf{i}=\sqrt{-1},\,\,\, j=0,1,2,\ldots n $$ Thus, we can
consider the function $F(\vec{x})$ which is defined on the convex
compact set $C_n=\Pi_2 \bigcap B[0,r_n]$ such that $F: C_n
\longmapsto C_n$, where $B[0,r_n]$ is a closed ball with radius
$r_n$. Function $F$ is continuous on the $C_n$ because of
$\vec{0}\not\in C_n$. Now we can use J.Schauder theorem.

\begin{theorem} [Schauder fixed-point theorem \cite{Schauder}]
Let $C$ be a compact convex subset of a normed space $E$. Then
each continuous map $F: C\longmapsto C$ has at least one fixed
point.
\end{theorem}

Hence, we can conclude that our system (\ref{functionF}) and
therefore system (\ref{system}) has at least one solution. Thus,
there is a function $\varphi(x)$ which satisfy to conditions
(\ref{conditions}) proposed lemma.

\begin{remark}
Let us a function $\varphi (x)$ satisfies lemma condition. If
$\beta\in \mathbf{R}$  then the function $\varphi (x+\beta)$ also
satisfies lemma condition. Moreover, if
$\displaystyle\int\limits_{-\infty}^{+\infty} \varphi^2 (x)\,
dx=\alpha $ then $\varphi (\alpha x)$ satisfies lemma condition.
\end{remark}

Thus, we can formulate the following result.

\begin{theorem}\label{theorem1}
For any integer $l$ there is a infinitely narrow soliton type
solution of the Hopf equation (in the sense of the definition
\ref{definition4}) up to $e^{-l}$ with respect to the norm
$|\cdot|_\nu$, i.e.
\begin{equation}
v(t,x,\varepsilon)=l_0+\Delta l \varphi
\left(\frac{x-ct}{\varepsilon}\right),
\end{equation}
$l_0, \Delta l, c$ are real numbers, $\Delta l\not=0$ and
$\varphi\in \mathcal{S}(\mathbf{R} )$,
$\displaystyle\int\limits_{-\infty}^{+\infty} \varphi(y)dy=1$.
Moreover,
\begin{equation}\label{ConstantConditions1}
\frac{c-l_0}{\Delta l}=\frac{1}{2}\int\limits_{-\infty}^{+\infty}
\varphi^2(x) d x.
\end{equation}
\end{theorem}

\begin{figure}[ht]
\begin{minipage}[b]{.49\linewidth}
\centering\includegraphics[width=\linewidth]{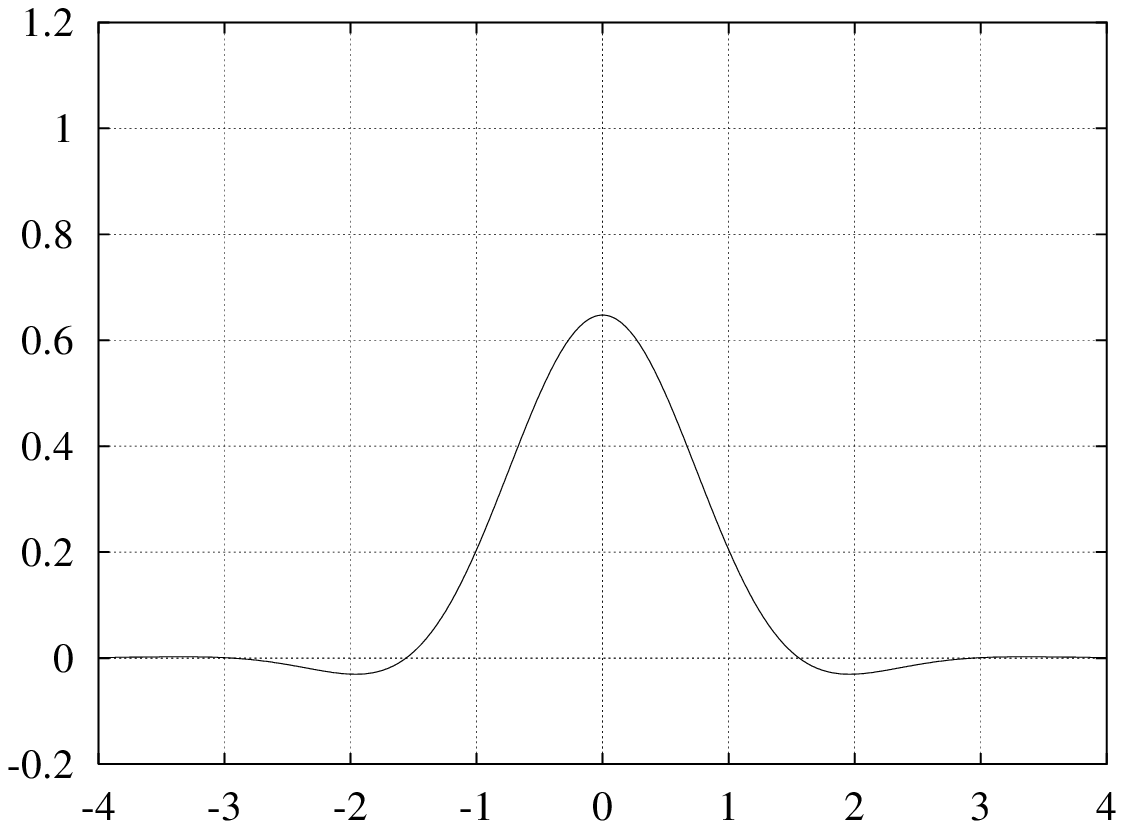}
\caption{\footnotesize The case $l=7$, $c=0.25032$.}\label{fi1}
\end{minipage}\hfill
\begin{minipage}[b]{.49\linewidth}
\centering\includegraphics[width=\linewidth]{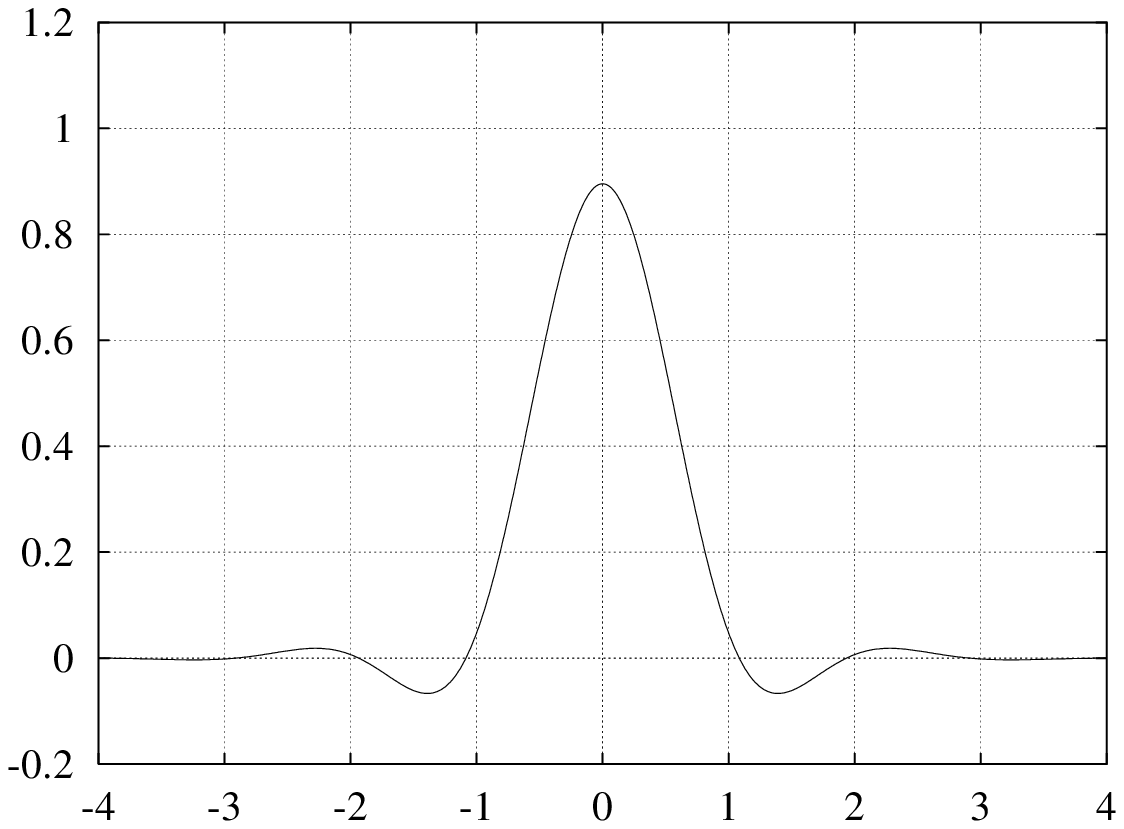}
\caption{\footnotesize The case $l=13$, $c=0.35442$.}\label{fi2}
\end{minipage}
\end{figure}

For example, calculations in case $l=7$ give the ``profile''
$\varphi (x)$ (see  Fig. \ref{fi1}) for the infinitely narrow
soliton $v(t,x,\varepsilon )=\varphi \left(\frac{x-c
t}{\varepsilon}\right)$:
\begin{equation}
\varphi(x)=\left\{ \frac{c_0}{\sqrt[4]{\pi}} +
\frac{c_2(4x^2-2)}{\sqrt{2^2 2!}\sqrt[4]{\pi}} +
\frac{c_4(16x^4-48x^2+12)}{\sqrt{2^4 4!}\sqrt[4]{\pi}}\right\}
e^{-x^2/2},
\end{equation}
where $c_0=0.66583$, $c_2=-0.23404$, $c_4=0.05028$, $c=0.25032$
($c$ is a velocity of the soliton). Numbers $c_0$, $c_2$, $c_4$
and $c$ were found approximately by iteration method using the
following sequence.
\begin{equation}
\vec{x}_{m+1}=A^{-1} (\mathcal{N}(\vec{x}_{m})), \,\,\,
m=0,1,2,\ldots .
\end{equation}
Matrix $A$ and a nonlinear $\mathcal{N}$ were introduced in the
lemma proof.

\begin{figure}[ht]
\begin{minipage}[b]{.49\linewidth}
\centering\includegraphics[width=\linewidth]{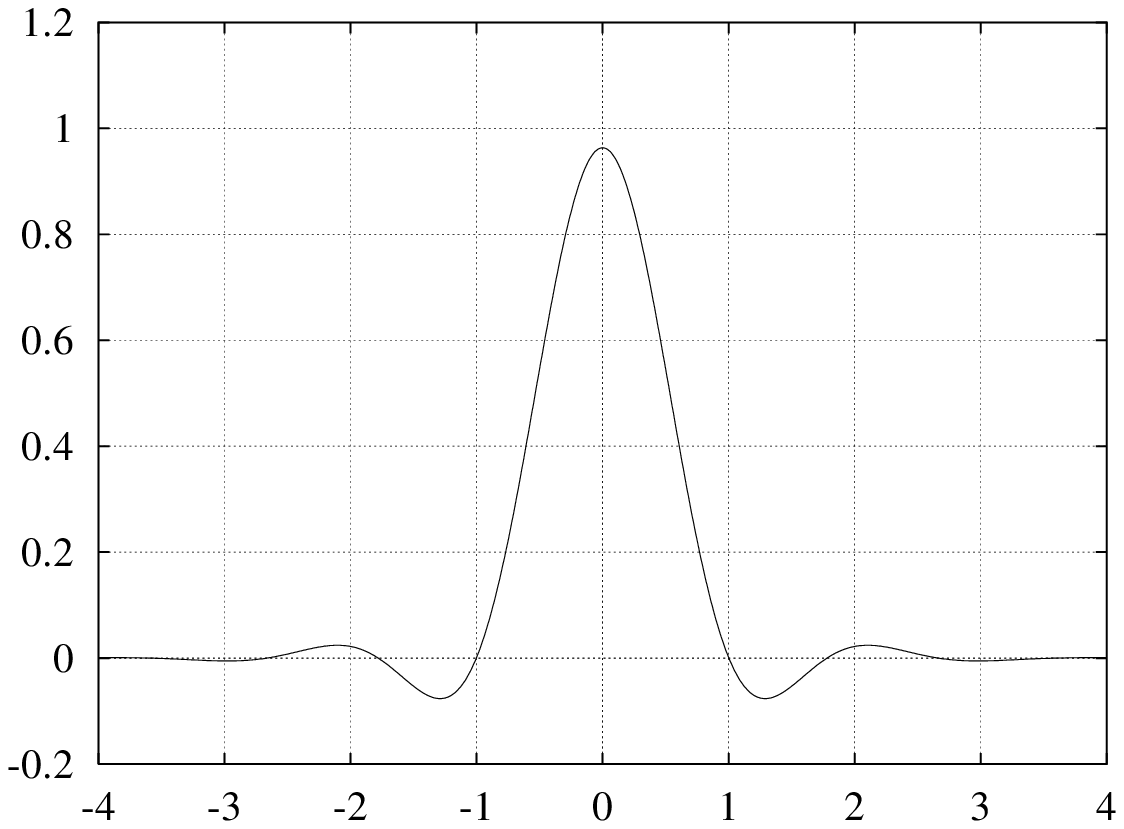}
\caption{\footnotesize The case $l=15$, $c=0.38267$.}\label{fi3}
\end{minipage}\hfill
\begin{minipage}[b]{.49\linewidth}
\centering\includegraphics[width=\linewidth]{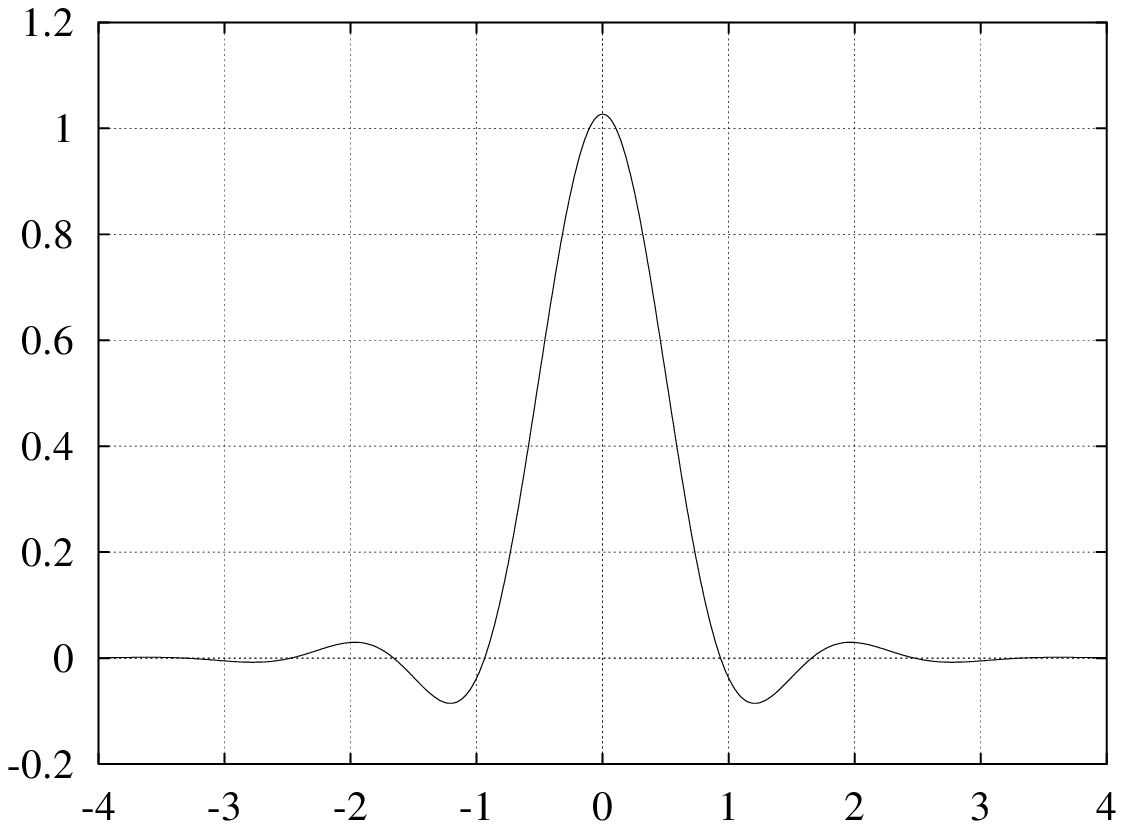}
\caption{\footnotesize The case $l=17$, $c=0.40892$.}\label{fi4}
\end{minipage}
\end{figure}

\begin{figure}[ht]
\begin{minipage}[b]{.49\linewidth}
\centering\includegraphics[width=\linewidth]{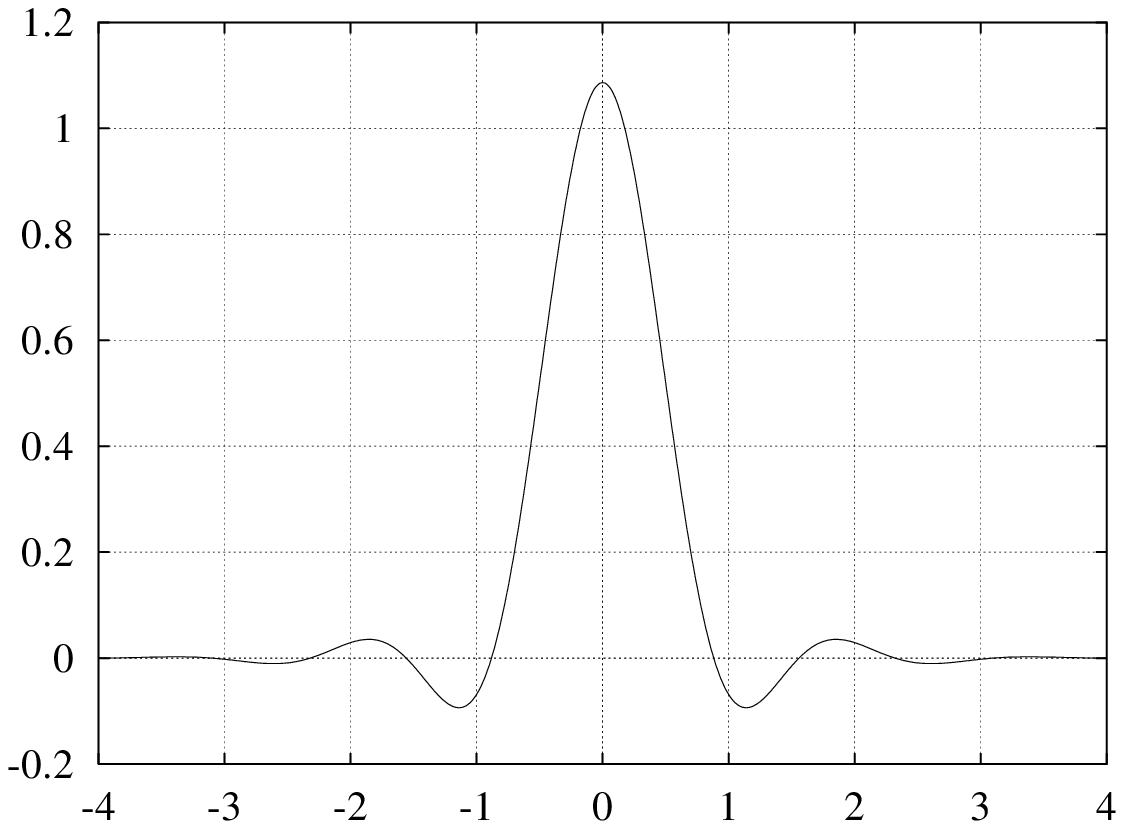}
\caption{\footnotesize The case $l=19$, $c=0.43357$.}\label{fi5}
\end{minipage}\hfill
\begin{minipage}[b]{.49\linewidth}
\centering\includegraphics[width=\linewidth]{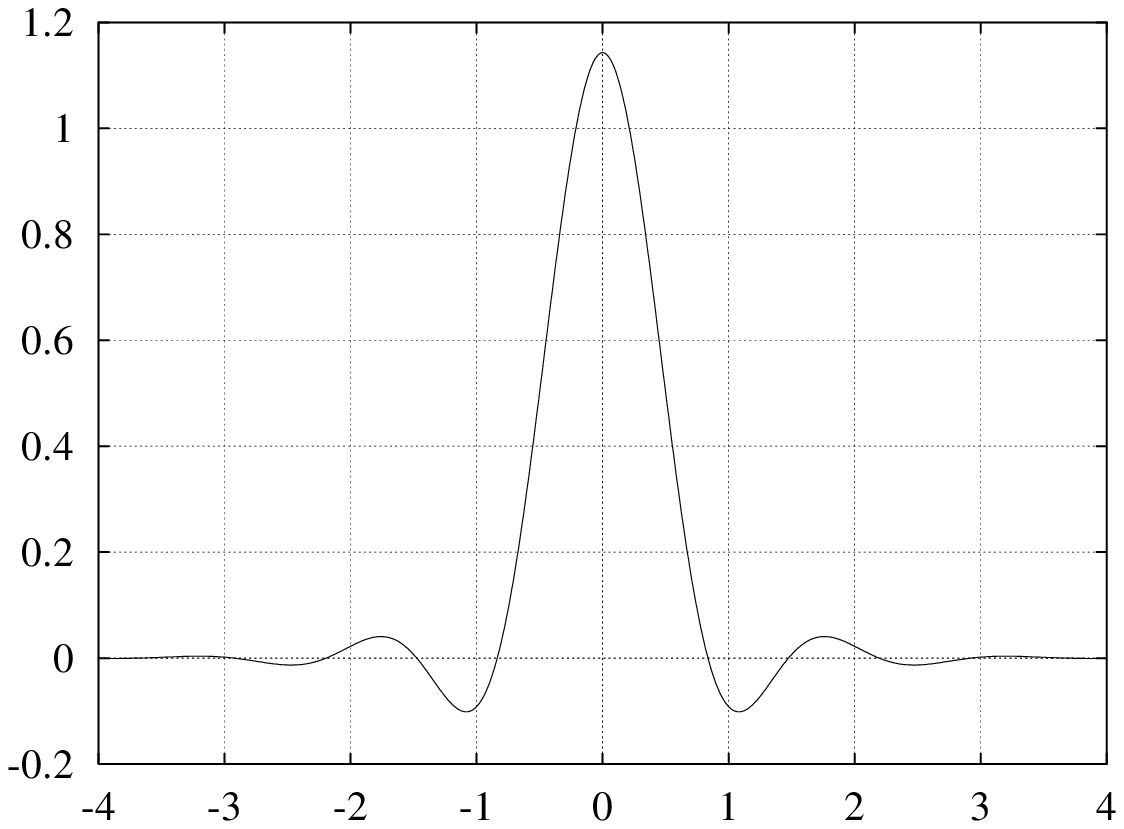}
\caption{\footnotesize The case $l=21$, $c=0.45678$.}\label{fi6}
\end{minipage}
\end{figure}

Calculations of soliton-like profiles $\varphi (x)$ for the Hopf
equation in case  $l=13,$ $15,$ $17,$ $19,$ $21$ give us pictures
(Fig. \ref{fi2}, \ref{fi3}, \ref{fi4}, \ref{fi5}, \ref{fi6}).

For the $l$ greater than $21$ matrix $A$ is close to singular and
calculations can be inaccurate.

\section{Calculations of the microscopic profiles of the
shock wave solutions of the Hopf equation in the sense of
$\mathbf{R}\langle\varepsilon\rangle$--distributions.}

A solution of the Hopf equation in this case we will seek in the
set $J$. Namely,

$$w(t,x,\varepsilon)=h_0+\Delta h K
\left(\frac{x-at}{\varepsilon}\right), $$ $h_0, \Delta h, a$ are
real numbers, $\Delta h\not=0$ and
$$K(x)=\int\limits_{-\infty}^{x} \theta (y) d y,
\int\limits_{-\infty}^{+\infty} \theta (y) d y=1, \quad \theta\in
\mathcal{S}(\mathbf{R} ).$$

Substitute $w(t,x,\varepsilon)$ into the integral expression
(\ref{solution-2}) using the following formulas

\begin{equation}
\int\limits_{-\infty}^{+\infty}\frac{\partial}{\partial t}\left\{K
\left(\displaystyle\frac{x-a t}{\varepsilon}\right)\right\}\psi
(x) dx=\sum\limits_{k=0}^{+\infty} (-a)\varepsilon^k m_k
\frac{\psi^{(k)}(at)}{k!},
\end{equation}
\begin{equation}
\int\limits_{-\infty}^{+\infty} K \left(\displaystyle\frac{x-a
t}{\varepsilon}\right)\frac{\partial}{\partial x}\left\{K
\left(\displaystyle\frac{x-a t}{\varepsilon}\right)\right\} \psi
(x) d x=\sum\limits_{k=0}^{+\infty} \varepsilon^k r_k
\frac{\psi^{(k)}(at)}{k!}.
\end{equation}

We denote by
\begin{equation}
m_k(\theta)=\int\limits_{-\infty}^{+\infty} y^k\theta(y) d
y,\,\,r_k(\theta)=\int\limits_{-\infty}^{+\infty} x^k \theta
(x)\left(\int\limits_{-\infty}^x \theta (y) dy\right) dx, \,\,
k=0,1,2,\ldots .
\end{equation}
Thus, we get
\begin{equation}\label{Laurent series Hopf}
\int\limits_{-\infty}^{+\infty}\left\{ w_t +ww_x\right\}\psi  d
x=\sum\limits_{k=0}^{+\infty} \left\{(\Delta h )^2 r_k-\Delta
h(a-h_0) m_k \right\}\varepsilon^{k} \frac{\psi^{(k)}(a t)}{k!}.
\end{equation}
From the last expression we have conditions for the function
$\theta(x)$
\begin{equation}\label{pre-conditions-Hopf}
r_k(\theta ) - \frac{a-h_0}{\Delta h} m_k(\theta )=0, \,\, k=0,1,2
\ldots.
\end{equation}
From the first ($k=0$) we have
\begin{equation}
\frac{a-h_0}{\Delta h}=\int\limits_{-\infty}^{+\infty} \theta (x)
\left(\int\limits_{-\infty}^x \theta (y) dy\right) dx=\frac{1}{2}.
\end{equation}
Therefore,  we can rewrite (\ref{pre-conditions-Hopf}) as
\begin{equation}\label{pre-conditions-1-Hopf}
\frac{1}{2}\int\limits_{-\infty}^{+\infty} x^k \theta(x)d
x=\int\limits_{-\infty}^{+\infty}x^k \theta (x)
\left(\int\limits_{-\infty}^x \theta (y) dy\right) dx \,\,k=0,1,2
\ldots.
\end{equation}
The same method one can prove that there is such function $\theta
(x)\in\mathcal{S}(\mathbf{R})$ which satisfies the following
conditions
\begin{equation}\label{conditions-Hopf}
\frac{1}{2}\int\limits_{-\infty}^{+\infty} x^k \theta(x)d
x=\int\limits_{-\infty}^{+\infty}x^k \theta (x)
\left(\int\limits_{-\infty}^x \theta (y) dy\right) dx \,\,k=0,1,2
\ldots n .
\end{equation}
Thus, we can formulate next result.
\begin{theorem}
For any integer $l$ there is a shock wave type solution of the
Hopf equation (in the sense of the definition \ref{definition4})
up to $e^{-l}$ with respect to the norm $|\cdot|_\nu$.
\begin{equation}
w(t,x,\varepsilon)=h_0+\Delta h K
\left(\frac{x-at}{\varepsilon}\right),
\end{equation}
$h_0, \Delta h, a$ are real numbers, $\Delta h\not=0$ and
$K(x)=\displaystyle\int\limits_{-\infty}^{x} \theta (y) d y$,
$\displaystyle\int\limits_{-\infty}^{+\infty} \theta (y) d y=1$
and $\theta\in \mathcal{S}(\mathbf{R} ).$ Moreover,
\begin{equation}\label{ConstantConditions2}
\frac{a-h_0}{\Delta h}=\frac{1}{2}.
\end{equation}
\end{theorem}

Note that the condition (\ref{ConstantConditions2}) is
\emph{Rankine --- Hugoniot condition} for the velocity of a shock
wave.

\begin{figure}[ht]
\begin{minipage}[b]{.49\linewidth}
\centering\includegraphics[width=\linewidth]{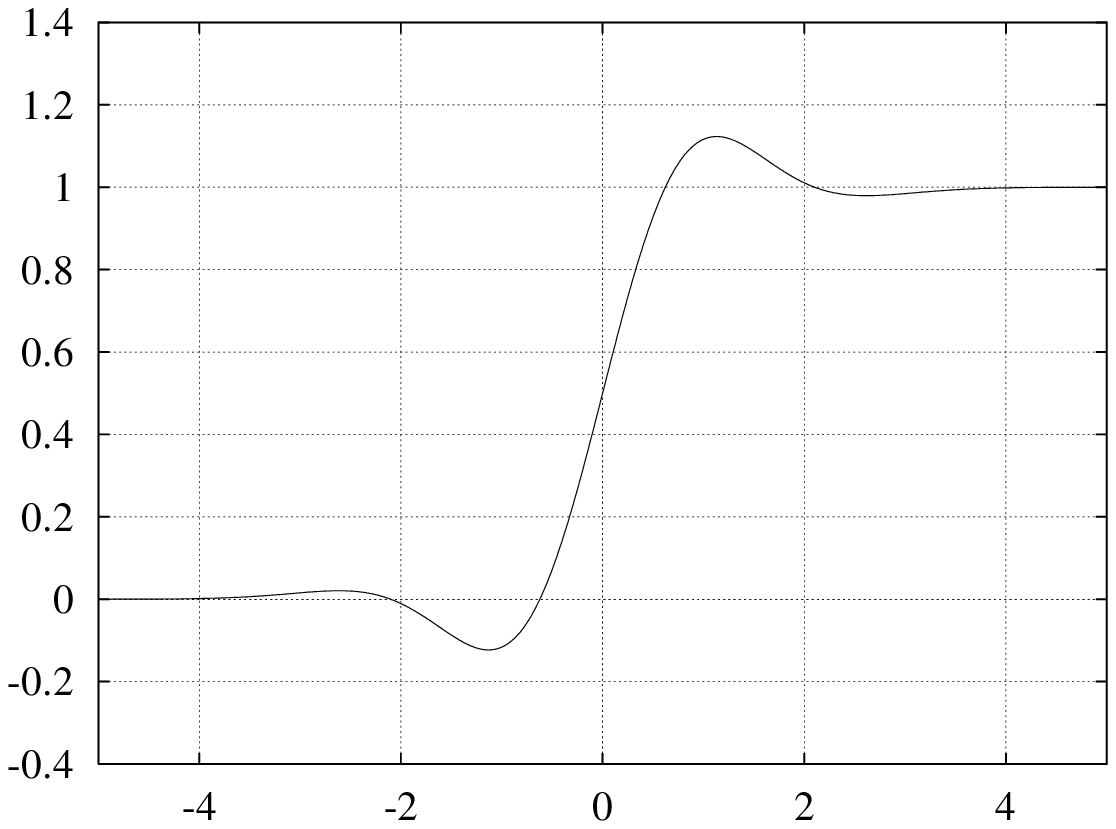}
\caption{\footnotesize Graph of the function $K(x)$.}\label{fi7}
\end{minipage}\hfill
\begin{minipage}[b]{.49\linewidth}
\centering\includegraphics[width=\linewidth]{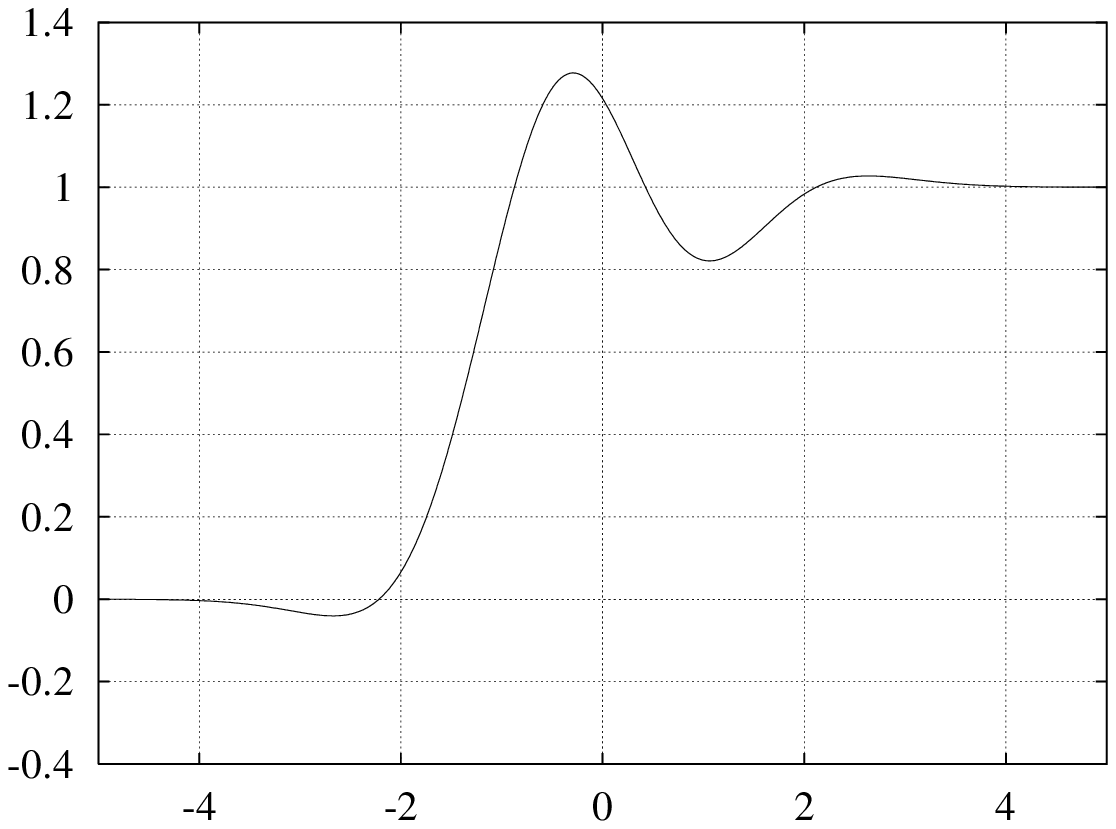}
\caption{\footnotesize Graph of the function $K_1(x)$.}\label{fi8}
\end{minipage}
\end{figure}

As in previous section   we  seek function  $\theta(x)$ in the
following type:
\begin{equation}
\varphi (x)=a_0 h_0(x)+a_1 h_1(x)+\ldots+a_{n} h_{n}(x),
\end{equation} where
$h_{k}(x)$ are Hermite functions. Calculations in case $l=7$ give
the following ``profile'' ($K(x)$) for the shock wave
$w(t,x,\varepsilon )=K\left(\displaystyle\frac{x-a
t}{\varepsilon}\right)$ (where $h_0=0, \Delta h=1$).

\begin{equation}\label{Firsttypeshock}
K(x)=\int\limits_{-\infty}^x \left\{ \frac{c_0}{\sqrt[4]{\pi}}
+\frac{c_2(4\tau^2-2)}{\sqrt{2^2
2!}\sqrt[4]{\pi}}+\frac{c_4(16\tau^4-48\tau^2+12)}{\sqrt{2^4
4!}\sqrt[4]{\pi}} \right\}e^{-\tau^2/2} d\tau
\end{equation}
where $c_0=0.79617$, $c_2=-0.53004$, $c_4=0.17923$, $c=1/2$ is a
velocity of the shock wave (see Fig. \ref{fi7}). Numbers $c_0$,
$c_2$, $c_4$ were found approximately.

Note that the function $K(x)$ is not unique. There is a different
function $K_1(x)$ which satisfies mentioned above conditions. It
has the following type
\begin{equation}\label{Secondtypeshock}
\begin{array}{l}
K_1(x)=\displaystyle\int\limits_{-\infty}^x
\left\{\frac{c_0}{\sqrt[4]{\pi}} +\frac{c_1 2\tau}{\sqrt{2^1
1!}\sqrt[4]{\pi}} +\frac{c_2 (4\tau^2-2)}{\sqrt{2^2
2!}\sqrt[4]{\pi}}\right\}e^{-\tau^2/2} d\tau +\\ \displaystyle
+\int\limits_{-\infty}^x\left\{\frac{c_3 (8 \tau^3-12\tau)
}{\sqrt{2^3
3!}\sqrt[4]{\pi}}+\frac{c_4(16\tau^4-48\tau^2+12)}{\sqrt{2^4
4!}\sqrt[4]{\pi}} \right\}e^{-\tau^2/2} d\tau
\end{array}
\end{equation}
where $c_0=0.18357$, $c_1=-0.73567$, $c_2=0.74733$, $c_3=0.15327$
$c_4=-0.29539$, $c=1/2$ is a velocity of the shock wave (see Fig.
\ref{fi8}). Coefficients  $c_0$, $c_1$, $c_2$, $c_3$, $c_4$ were
found approximately by the Newton iteration method.

\begin{figure}[ht]
\begin{minipage}[b]{.49\linewidth}
\centering\includegraphics[width=\linewidth]{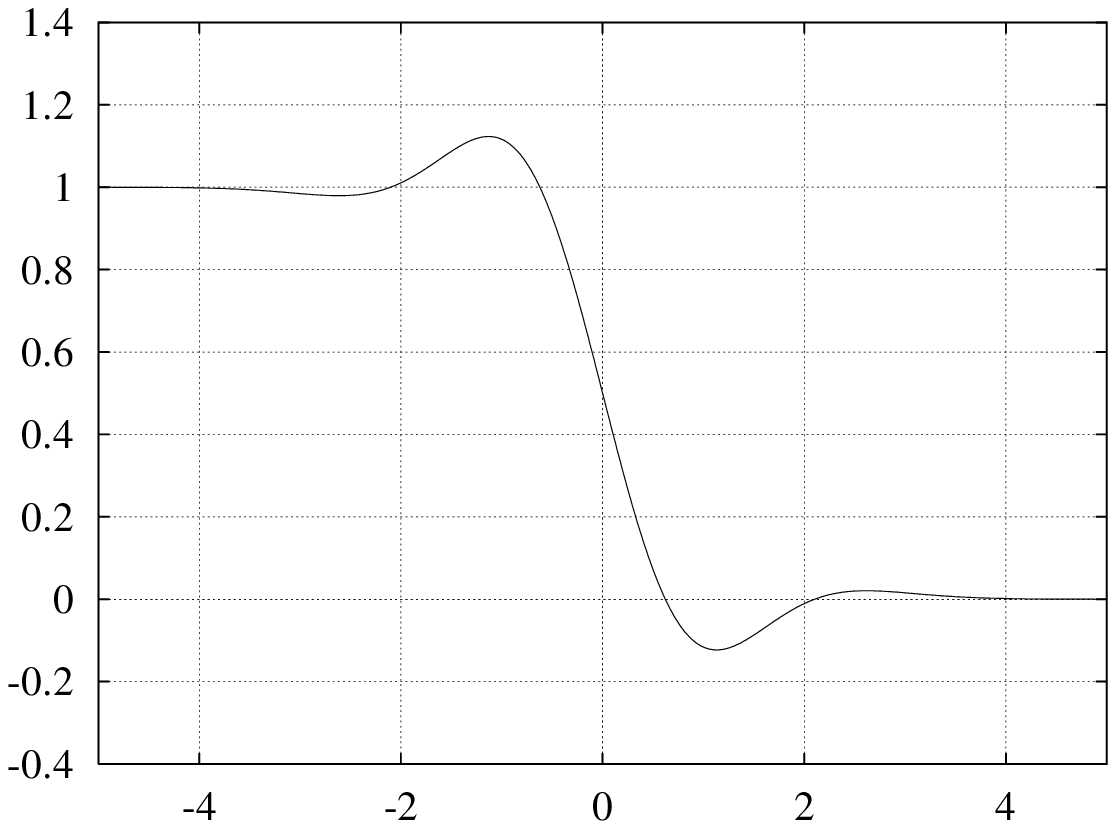}
\caption{\footnotesize First shock profile $1-K(x).$}\label{fi9}
\end{minipage}\hfill
\begin{minipage}[b]{.49\linewidth}
\centering\includegraphics[width=\linewidth]{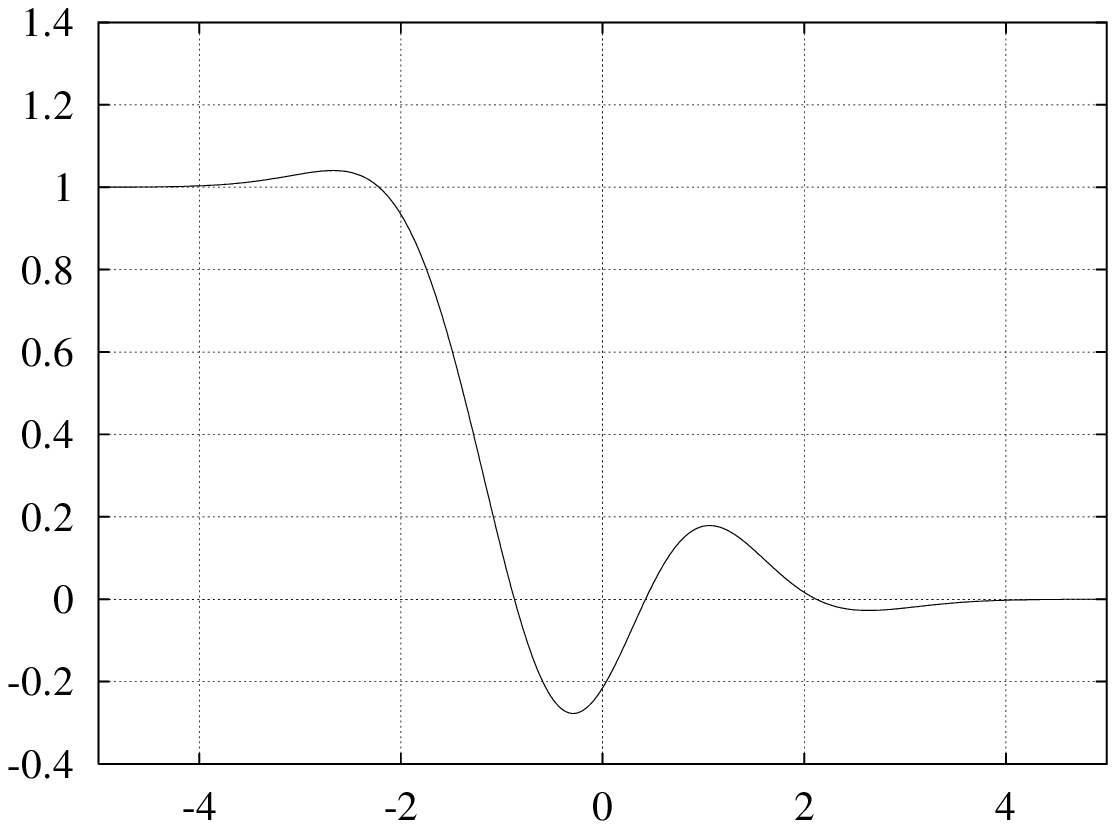}
\caption{\footnotesize Second shock profile
$1-K_1(x).$}\label{fi10}
\end{minipage}
\end{figure}

Taking in account the \emph{Rankine
--- Hugoniot condition} (\ref{ConstantConditions2}) we also have
graphs (Fig. \ref{fi9}, \ref{fi10}) as a shock profiles.

Here we describe how it is possible to find coefficients $c_0,
c_1, \ldots , c_n$  in this case by the Newton iteration method
for the following system of nonlinear equations.
\begin{equation}
P(\vec{c})=A\vec{c} -2\sum\limits_{k=0}^n
(S(k)\vec{c},\vec{c})\vec{e}_k =0, \,\,\, \vec{c}=(c_0, c_1,
\ldots , c_n)
\end{equation}
Vector $\vec{e}_k=(e_0, e_1, \ldots , e_n)$ such that $e_k=1$ and
$e_j=0$ for all $j\neq k$. $S(k)$ are matrices with elements
\begin{equation}
S_{ij}(k)=\int\limits_{-\infty}^{+\infty} x^k
h_i(x)\cdot\int\limits_{-\infty}^{x} h_j(y)\, d y\, d x,\,\,\,
i,j,k=0,1,2 \ldots n
\end{equation}

Matrix $A$ have elements
\begin{equation}
A_{ij}=\int\limits_{-\infty}^{+\infty} x^i h_j(x)\, d x ,\,\,\,
i,j=0,1,2 \ldots n
\end{equation}
We can write the formula for the Newton iteration method
\cite{Kantorovich-Akilov}.

\begin{equation}
\vec{x}_{m+1}=\vec{x}_{m}-\left[P'(\vec{x}_{m})\right]^{-1}\left[P(\vec{x}_{m})\right],
\end{equation}
where $\left[P'(\vec{x})\right]$ is a linear map depending on the
vector $\vec{x}$.
\begin{equation}
\left[P'(\vec{x})\right] [\vec{h}]=A\vec{h}
-2\left\{\sum\limits_{k=0}^n (S(k)\vec{x},\vec{h})\vec{e}_k
+\sum\limits_{k=0}^n (S^{T}(k)\vec{x},\vec{h})\vec{e}_k \right\}
\end{equation}

\begin{figure}[ht]
\begin{minipage}[b]{.49\linewidth}
\centering\includegraphics[width=\linewidth]{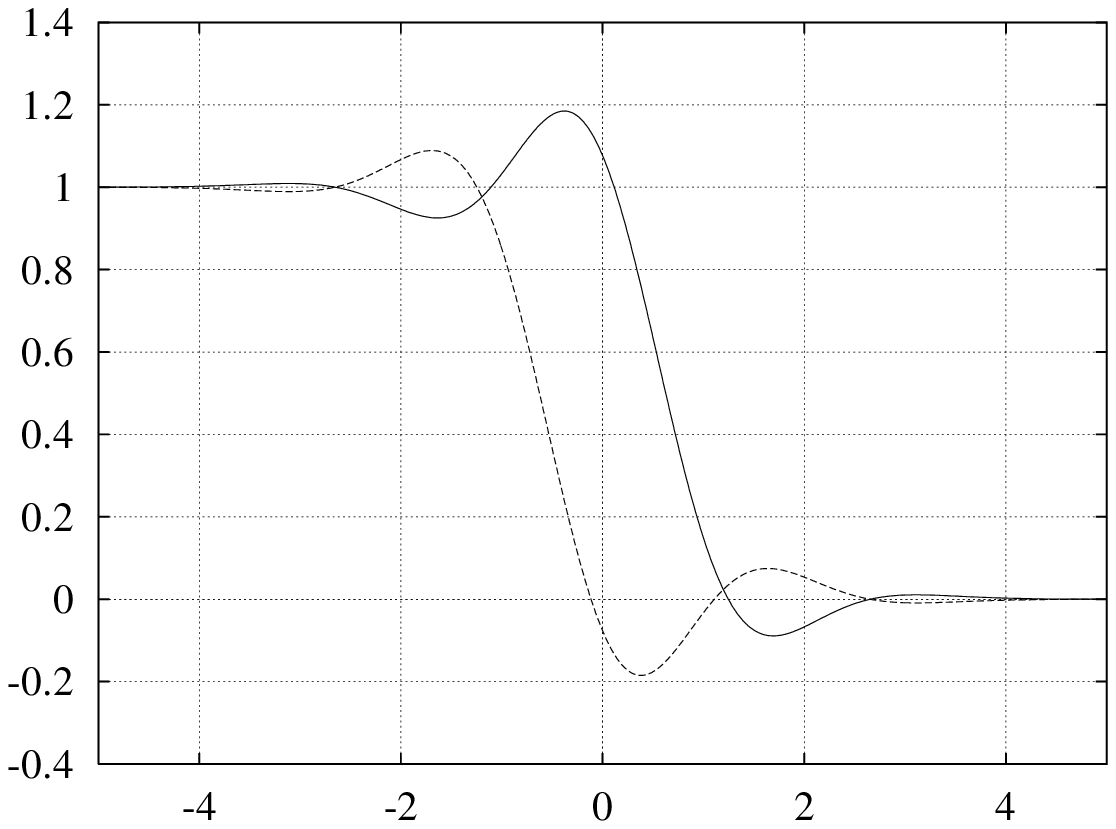}
\caption{\footnotesize Shock profiles when $l=8.$  }\label{fi11}
\end{minipage}\hfill
\begin{minipage}[b]{.49\linewidth}
\centering\includegraphics[width=\linewidth]{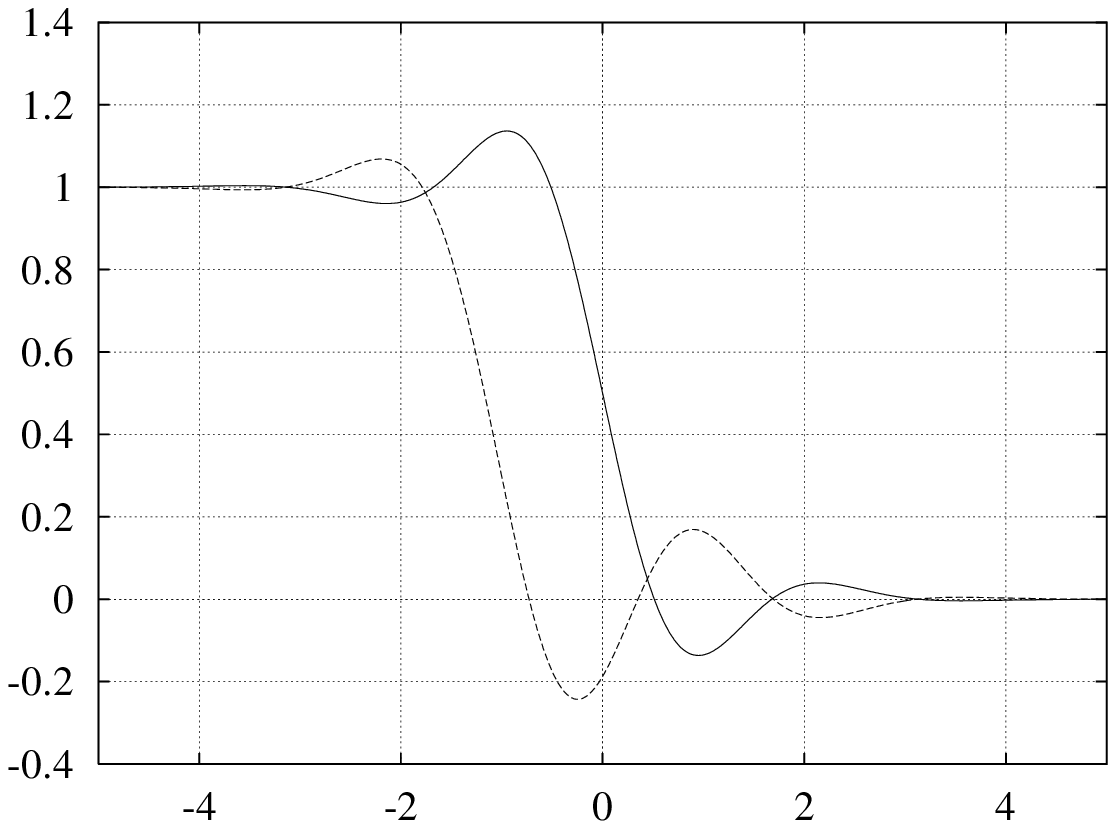}
\caption{\footnotesize Shock profiles when $l=9.$}\label{fi12}
\end{minipage}
\end{figure}

\begin{figure}[ht]
\begin{minipage}[b]{.49\linewidth}
\centering\includegraphics[width=\linewidth]{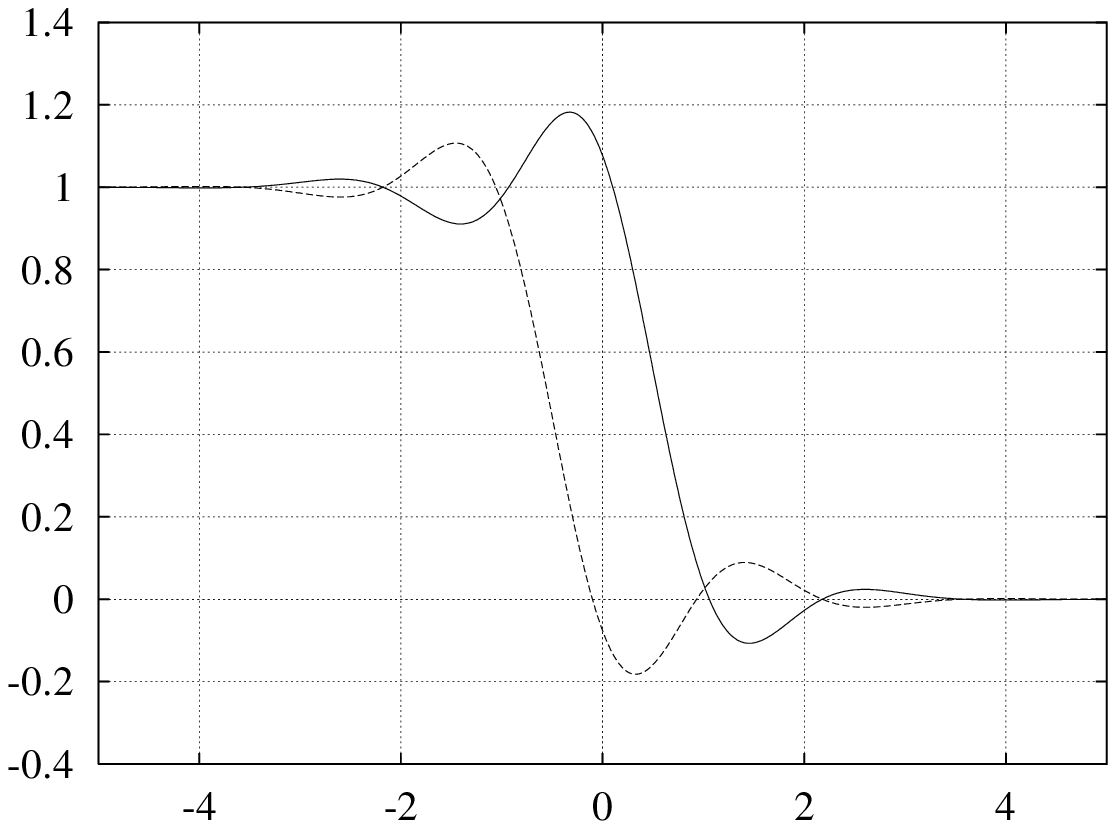}
\caption{\footnotesize Shock profiles when $l=10.$  }\label{fi13}
\end{minipage}\hfill
\begin{minipage}[b]{.49\linewidth}
\centering\includegraphics[width=\linewidth]{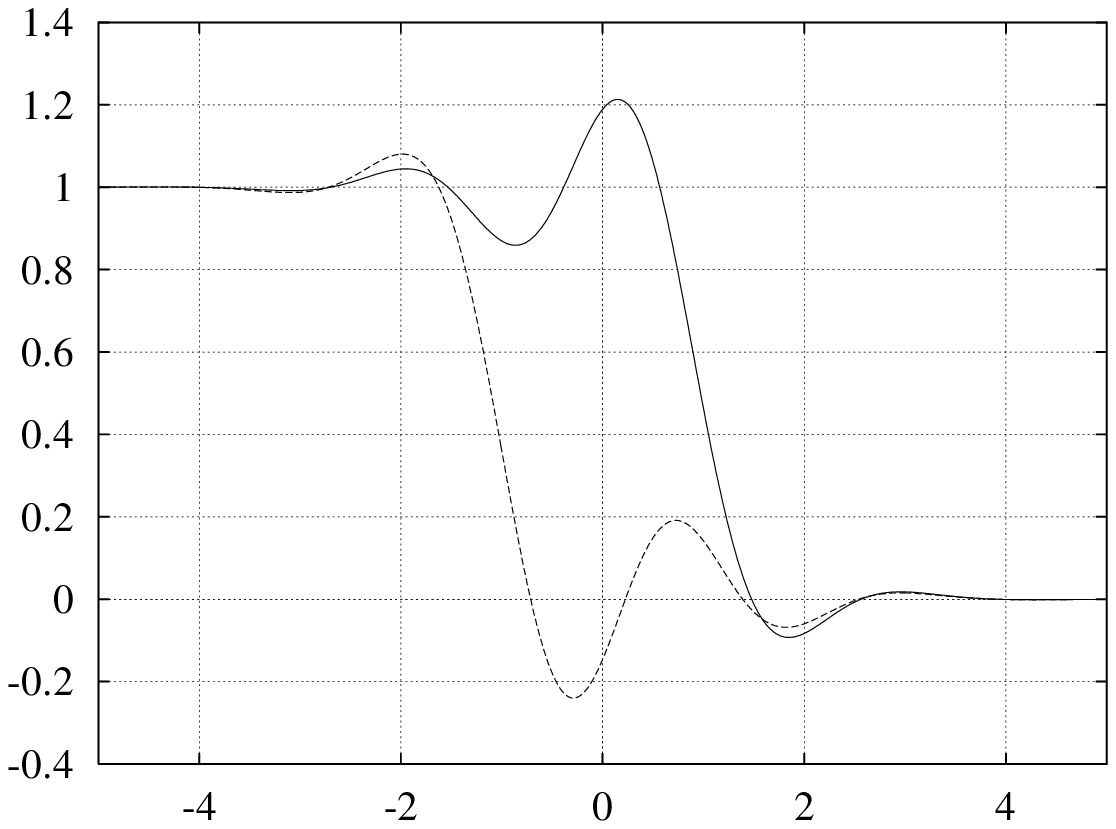}
\caption{\footnotesize Shock profiles when $l=11.$}\label{fi14}
\end{minipage}
\end{figure}

\begin{figure}[ht]
\begin{minipage}[b]{.49\linewidth}
\centering\includegraphics[width=\linewidth]{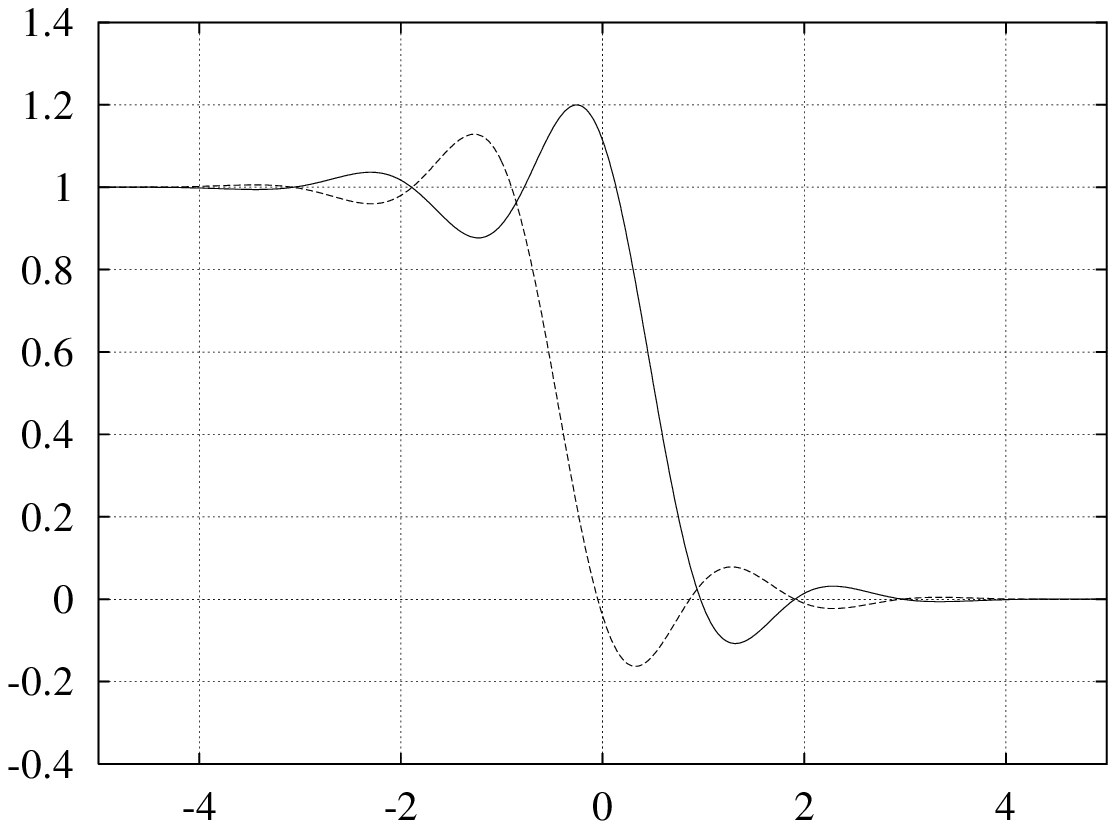}
\caption{\footnotesize Shock profiles when $l=12.$ }\label{fi15}
\end{minipage}\hfill
\begin{minipage}[b]{.49\linewidth}
\centering\includegraphics[width=\linewidth]{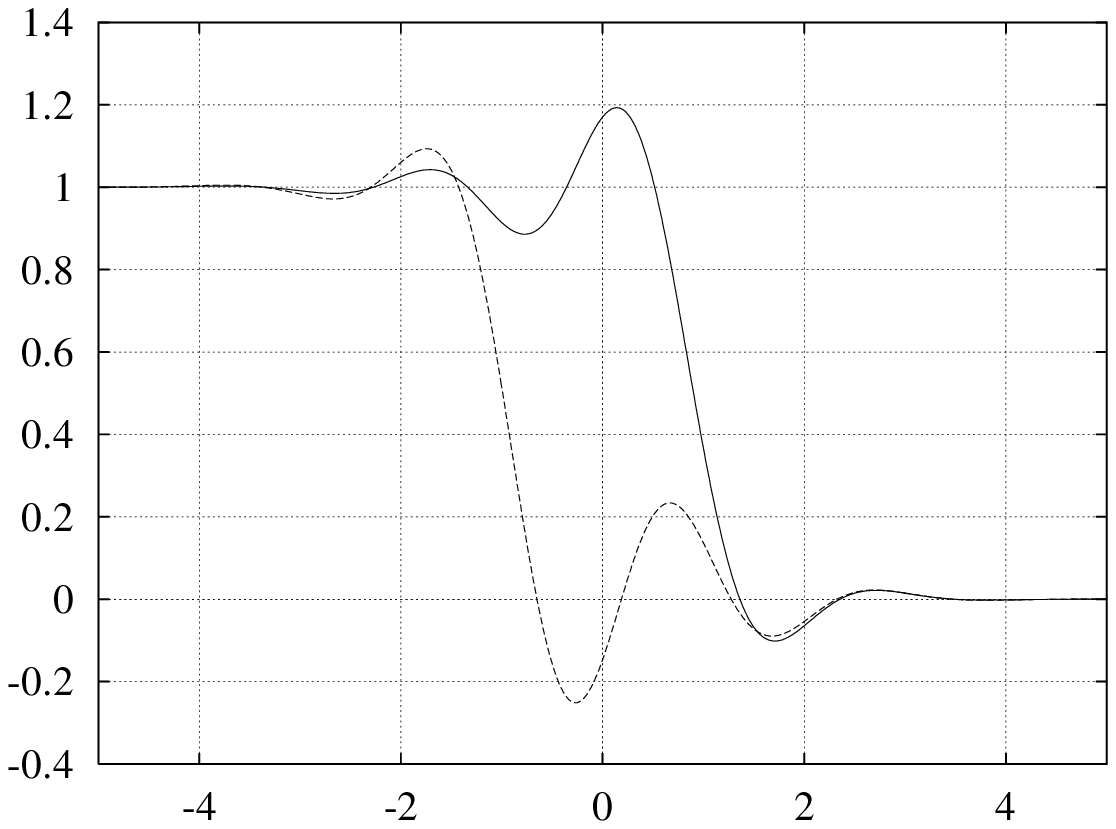}
\caption{\footnotesize Shock profiles when $l=13.$}\label{fi16}
\end{minipage}
\end{figure}

Calculations of shock profiles $K(x)$ for the Hopf equation in
case  $l=8,$ $9,$ $10,$ $11,$ $12,$ $13$ give us the following
pictures (Fig. \ref{fi11}, \ref{fi12}, \ref{fi13}, \ref{fi14},
\ref{fi15}, \ref{fi16}). Here, we show only two different types of
the shock type solutions of the Hopf equation. We can find more
solutions if we take a different initial data for the Newton
iteration method.

\begin{remark}
It is not easy to see  that there is exist function
\begin{equation}
\theta(x)=\sum\limits_{n=1}^{\infty} a_n h_n(x),\,\,\,
\vec{a}=(a_0, a_1, \ldots , a_n ,\ldots)\in l_2,
\end{equation}
such that
\begin{equation}
\frac{1}{2}\int\limits_{-\infty}^{+\infty} x^k \theta(x)d
x=\int\limits_{-\infty}^{+\infty}x^k \theta (x)
\left(\int\limits_{-\infty}^x \theta (y) dy\right) dx, \,\,k=0,1,2
\ldots.
\end{equation}
We think that it is true.
\end{remark}

\section{Calculations of the microscopic profiles of the shock wave
solutions of the  equations of compressible flow in the sense of
$\mathbf{R}\langle\varepsilon\rangle$--distributions.}

Now, we are going to study the equations of compressible flow
\begin{equation}\label{momentum1}
u_t+p_x=0
\end{equation}
\begin{equation}\label{mass1}
v_t-u_x=0
\end{equation}
\begin{equation}\label{work'1}
\frac{1}{\gamma -1} \left(pv \right)_t+pu_x=0
\end{equation}
in specific sense. Namely, we rewrite mentioned equations in the
sense of $\mathbf{R}\langle\varepsilon\rangle$--distributions.

We will seek for a solution of the equations  in the following
form
\begin{equation}\label{velocity-function}
u(t,x,\varepsilon)=u_0+\Delta u U
\left(\frac{x-at}{\varepsilon}\right),
\end{equation}
$u_0, \Delta u, v$ are real numbers, $\Delta u\not=0$ and
$U(x)=\displaystyle\int\limits_{-\infty}^{x} \widetilde{U} (y) d
y$, $\displaystyle\int\limits_{-\infty}^{+\infty} \widetilde{U}
(y) d y=1$ and $\widetilde{U}\in \mathcal{S}(\mathbf{R} ).$
\begin{equation}\label{pressure-function}
p(t,x,\varepsilon)=p_0+\Delta p P
\left(\frac{x-at}{\varepsilon}\right),
\end{equation}
$p_0, \Delta p, a$ are real numbers, $\Delta p\not=0$ and
$P(x)=\displaystyle\int\limits_{-\infty}^{x} \widetilde{P} (y) d
y$, $\displaystyle\int\limits_{-\infty}^{+\infty} \widetilde{P}
(y) d y=1$ and $\widetilde{P}\in \mathcal{S}(\mathbf{R} ).$
\begin{equation}\label{volume-function}
v(t,x,\varepsilon)=v_0+\Delta v V
\left(\frac{x-at}{\varepsilon}\right),
\end{equation}
$v_0, \Delta v, a$ are real numbers, $\Delta v\not=0$ and
$V(x)=\displaystyle\int\limits_{-\infty}^{x} \widetilde{V} (y) d
y$, $\displaystyle\int\limits_{-\infty}^{+\infty} \widetilde{V}
(y) d y=1$ and $\widetilde{V}\in \mathcal{S}(\mathbf{R} ).$ Note
that $a$ is a velocity of the shock waves.

In the other hand, we suppose
\begin{equation}\label{velocity-expansion}
\widetilde{U}(x)=a_0 h_0(x)+a_1 h_1(x)+\ldots+a_{n} h_{n}(x),\quad
\vec{a}=(a_0, a_1, \ldots , a_n),
\end{equation}
\begin{equation}\label{pressure-expansion}
\widetilde{P}(x)=b_0 h_0(x)+b_1 h_1(x)+\ldots+b_{n} h_{n}(x),\quad
\vec{b}=(b_0, b_1, \ldots , b_n),
\end{equation}
\begin{equation}\label{volume-expansion}
\widetilde{V}(x)=c_0 h_0(x)+c_1 h_1(x)+\ldots+c_{n} h_{n}(x),\quad
\vec{c}=(c_0, c_1, \ldots , c_n),
\end{equation}
where $h_{k}(x)$ are Hermite functions.

We understand the  solution of the system in sense of
$\mathbf{R}\langle\varepsilon\rangle$--distributions.
\begin{definition}\label{definition5}
Functions $u\in J$, $p\in J$ and $v\in J$ is a solution of the
system (\ref{momentum1}), (\ref{mass1}), (\ref{work'1}) up to
$e^{-l}$, $l\in \mathbf{N}_0$ in the sense of
$\mathbf{R}\langle\varepsilon\rangle$--distributions if for any
$t\in [0,T]$
\begin{equation}\label{solution-1ofthesystem-1}
\int\limits_{-\infty}^{+\infty} \left\{u_t+ p_x\right\}\psi (x)
dx=\sum\limits_{k=l}^{+\infty} \xi_k \varepsilon^k\in
\mathbf{R}\langle\varepsilon\rangle,
\end{equation}
\begin{equation}\label{solution-2ofthesystem-1}
\int\limits_{-\infty}^{+\infty}\left\{v_t- u_x\right\}\psi (x)
dx=\sum\limits_{k=l}^{+\infty} \zeta_k \varepsilon^k\in
\mathbf{R}\langle\varepsilon\rangle
\end{equation}
\begin{equation}\label{solution-3ofthesystem-1}
\int\limits_{-\infty}^{+\infty}\left\{\frac{1}{\gamma
-1}\left\{p_t v+p v_t\right\} +p_t u_x\right\}\psi (x)
dx=\sum\limits_{k=l}^{+\infty} \eta_k \varepsilon^k\in
\mathbf{R}\langle\varepsilon\rangle
\end{equation}
for every $\psi\in\mathcal{S}(\mathbf{R} )$.

In case when $l$ is equal to  $+\infty$  functions $u
(t,x,\varepsilon )$,  $p(t,x,\varepsilon )$ and $v(t,x,\varepsilon
)$ exactly satisfies the system (\ref{momentum1}), (\ref{mass1}),
(\ref{work'1}) in the sense of
$\mathbf{R}\langle\varepsilon\rangle$--distributions.
\end{definition}

Substituting $u$, $p$ and $v$ into
(\ref{solution-1ofthesystem-1}), (\ref{solution-2ofthesystem-1})
we get the following relations for the moments.
\begin{equation}\label{system-1conditions-1}
-a \Delta u m_{k}(\widetilde{U}) + \Delta p m_k(\widetilde{P})=0,
\,\,\,k=0,1,2,\ldots n,
\end{equation}
\begin{equation}\label{system-1conditions-2}
-a \Delta v m_{k}(\widetilde{V}) - \Delta u m_k(\widetilde{U})=0,
\,\,\,k=0,1,2,\ldots n.
\end{equation}

We denote as usual by
\begin{equation}
m_k(\widetilde{U})=\int\limits_{-\infty}^{+\infty} x^k
\widetilde{U}(x) dx, \,\,
m_k(\widetilde{U}U)=\int\limits_{-\infty}^{+\infty} x^k
\widetilde{U} (x)\left(\int\limits_{-\infty}^x \widetilde{U} (y)
dy\right) dx,
\end{equation}
where $k=0,1,2,\ldots .$

It is easy to find $a$ from (\ref{system-1conditions-1}) and
(\ref{system-1conditions-1}) when k=0. Indeed,
\begin{equation} \label{Hugoniot1-2}
a=\frac{\Delta p}{\Delta u}, \,\, a=-\frac{\Delta u}{\Delta v}
\end{equation}
These are  \emph{Rankine
--- Hugoniot conditions.} Indeed, from the last
expressions we have
\begin{equation}\label{Hugoniot12}
a^2=-\frac{\Delta p}{\Delta v}.
\end{equation}
See J. von Neumann and R.D. Richtmyer \cite{Neumann-Richtmyer}
formula (21). We also conclude that $\Delta u<0$, $\Delta p<0$ and
$\Delta v >0$.

Because of $\Delta u$, $\Delta p$ and $\Delta v$ some real
numbers, therefore, all three vectors with coordinates
$m_{k}(\widetilde{U})$, $m_{k}(\widetilde{P})$ and
$m_k(\widetilde{V})$, $k=0,1,2,\ldots n$, respectively should be
collinear. However,
$$m_{0}(\widetilde{U})=m_{0}(\widetilde{P})=m_{0}(\widetilde{P})=1.$$
Hence, $\vec{a}=\vec{b}=\vec{c}$.

Substituting $u$, $p$ and $v$ into (\ref{solution-3ofthesystem-1})
and taking in account the last equalities  we get the following
relations for the moments $(k=0,1,\ldots n):$
\begin{equation}\label{system-1conditions-3}
\left\{p_0\Delta u - \frac{a(\Delta p v_0 +\Delta v p_0)}{\gamma
-1}  \right\} m_{k}(\widetilde{U}) + \left\{\Delta p \Delta u -
\frac{2a\Delta p \Delta v}{\gamma -1}  \right\}
m_{k}(U\widetilde{U}) =0.
\end{equation}
When $k=0$ we will have \emph{Rankine
--- Hugoniot conditions} for our system
\begin{equation}\label{Hugoniot3}
\left\{p_0\Delta u - \frac{a(\Delta p v_0 +\Delta v p_0)}{\gamma
-1}  \right\} + \frac{1}{2}\left\{\Delta p \Delta u -
\frac{2a\Delta p \Delta v}{\gamma -1}  \right\}=0.
\end{equation}
Using the equalities (\ref{Hugoniot1-2}), (\ref{Hugoniot3}), we
will have
\begin{equation}\label{H-adiabada}
\gamma \cdot p_0 + \frac{\Delta p}{\Delta v}\cdot v_0
+\frac{1}{2}\cdot (\gamma +1) \cdot \Delta p=0 \,\,\,\textrm{or}
\,\,\, \frac{p_1}{p_0}=\frac{(\gamma +1) v_0 -(\gamma
-1)v_1}{(\gamma +1) v_1 -(\gamma -1)v_0}.
\end{equation}
Then taking in account the condition (\ref{Hugoniot12}), we get
\begin{equation}
\left\{v_0 + \frac{1}{2}\cdot (\gamma +1)\Delta v\right\}
a^2=\gamma p_0.
\end{equation}
Hence,
\begin{equation}\label{formula}
a=\left(\frac{2}{(\gamma +1)\cdot\frac{\Delta v}{v_0} +
2}\right)^{1/2} \left(\gamma\cdot \frac{p_0}{v_0}\right)^{1/2}.
\end{equation}
Thus, we get the same formula for the shock velocity as in the
paper of J. von Neumann and R.D. Richtmyer
\cite{Neumann-Richtmyer} (formula (72)).

Thus, from (\ref{system-1conditions-3}) follows that
$$m_{k}(\widetilde{U})=2m_{k}(\widetilde{U}U), \,\,\,
k=0,1,2,\ldots n.$$ This system we already know how to solve by
the Newton iteration method. See conditions
(\ref{conditions-Hopf}) and solutions  in this case. Thus, we can
formulate the following result.
\begin{theorem}
For any integer $l$ there is a solution of the system of equations
(\ref{momentum1}), (\ref{mass1}), (\ref{work'1}) in the sense of
the definition \ref{definition5} if $$ a=\frac{\Delta p}{\Delta
u}, \,\, a=-\frac{\Delta u}{\Delta v}, $$ $$\left\{p_0\Delta u -
\frac{a}{\gamma -1} (\Delta p v_0 +\Delta v p_0) \right\} +
\frac{1}{2}\left\{\Delta p \Delta u - \frac{2a}{\gamma -1} \Delta
p \Delta v \right\}=0. $$
\end{theorem}
\begin{figure}[ht]
\begin{minipage}[b]{.49\linewidth}
\centering\includegraphics[width=\linewidth]{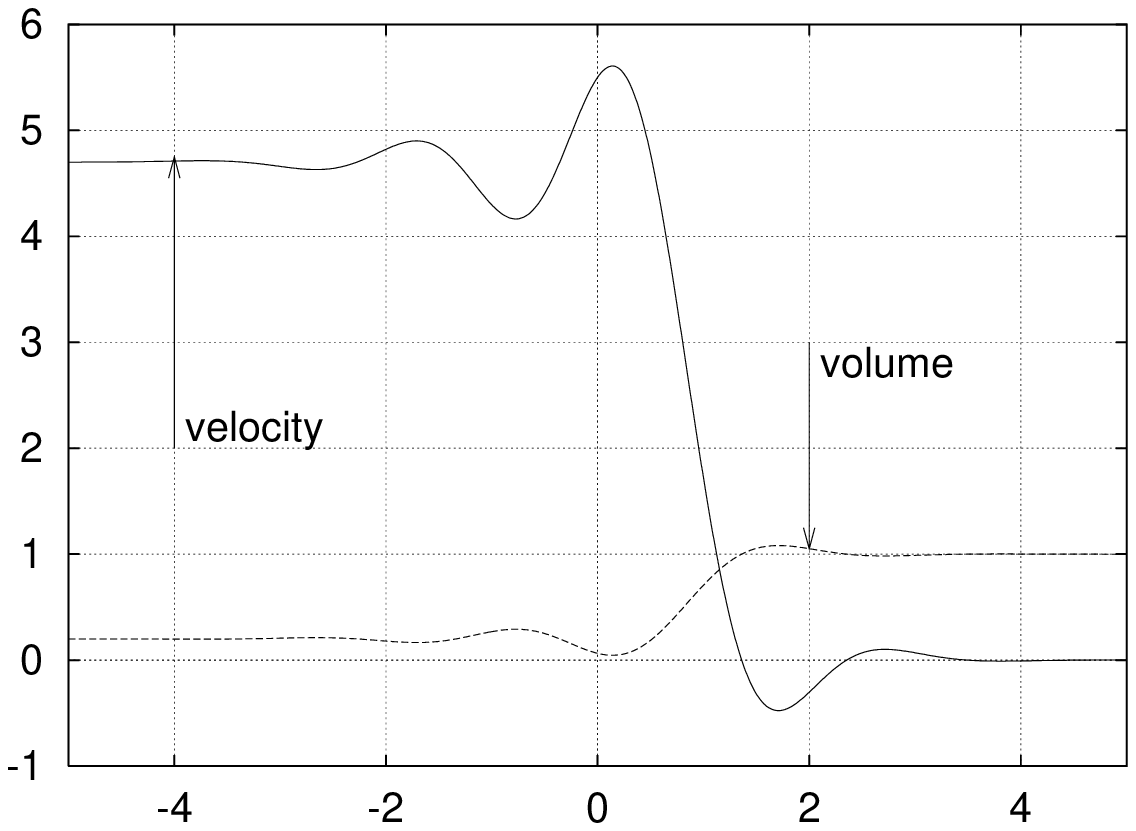}
\caption{\footnotesize Velocity and volume.}\label{fi17}
\end{minipage}\hfill
\begin{minipage}[b]{.49\linewidth}
\centering\includegraphics[width=\linewidth]{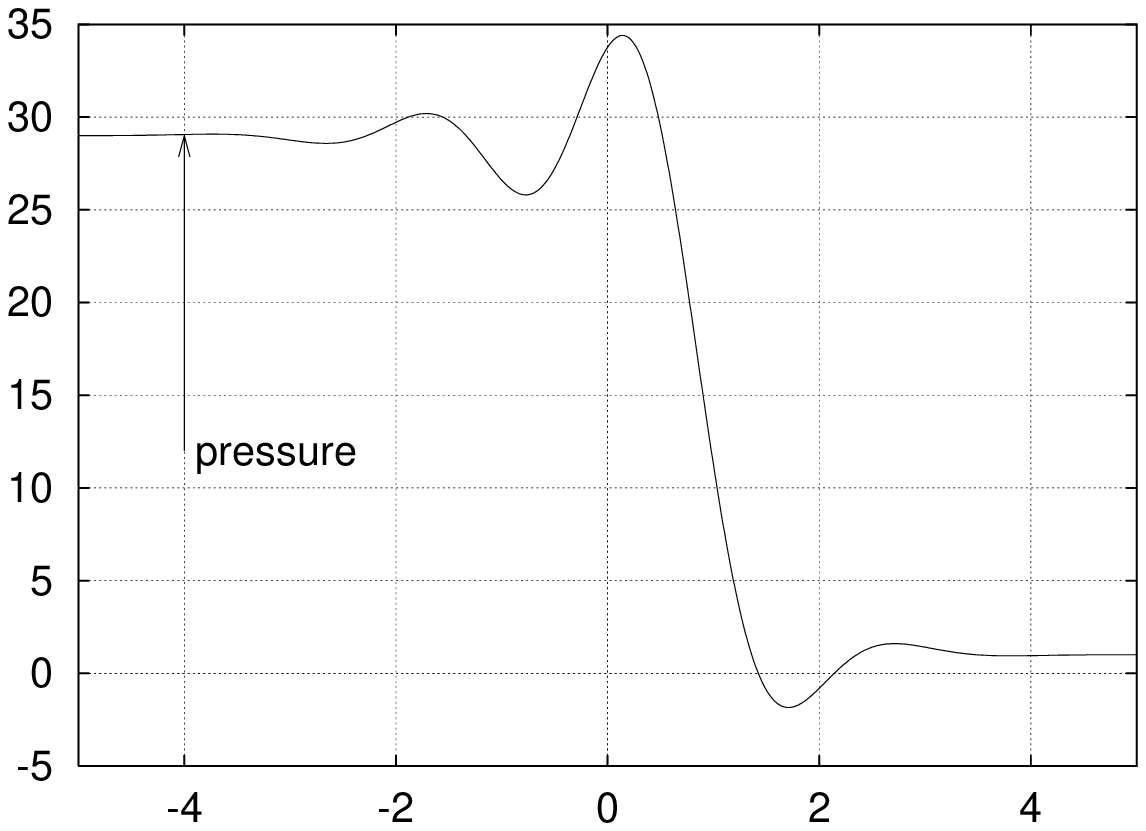}
\caption{\footnotesize Shock profile of pressure.}\label{fi18}
\end{minipage}
\end{figure}

If we assume that $v_0$ and $\Delta v$ are known then we can
calculate all constants $\Delta u$, $\Delta p$,  $a$ and profiles
of the shocks. Let us take $v_0=0.2$, $\Delta v=0.8$. We denote by
$v_1=v_0+\Delta v$, $v_1$ is a volume before the shock formation.
The quantity $\eta =v_1/v_0$ is a measure of the shock strength.
Denote by $p_1=p_0+\Delta p$, $v_1$ is a normal pressure (it is
known, $p_1 =1\,\, atm$) before the shock formation. From the
(\ref{H-adiabada}) we can get $$\Delta p=\frac{2\gamma p_1}{\gamma
-2v_0/\Delta v-1}.$$ Therefore, we can calculate step by step
$$a=\sqrt{-\frac{\Delta p}{\Delta v}}, \,\, \Delta u
=-a\cdot\Delta v.$$

Shock profiles of the considered system (\ref{momentum1}),
(\ref{mass1}), (\ref{work'1}) one can find on pictures (Fig.
\ref{fi17}, \ref{fi18}). We considered only one type profile of
the shock, the case when $l=13$, $v_0=0.2$, $\Delta v=0.8$, $p_1
=1$ and $\gamma =1.4$.

Finally, $\Delta p =-28p_1$, $a=5.9$, $\Delta u =-4.7$,
$\vec{a}=(0.47894,$ $0.70727,$ $0.31783,$ $-0.4706,$ $-0.53605,$ \
\ $0.20792,$ \ \ $0.4099,$\ \ $-0.057108,$ \ \ $-0.16832,$
$0.0088166,$ $0.029395)$.

In conclusion we should emphasis that our calculation method looks
like the Fourier method for linear differential equations but
applied to the nonlinear equations. Our method allowed to obtain
all known formulas for the shocks characteristics and, in
addition, find a microscopic behaviour of shocks in the thin
layer. According to the our model the pressure in the thin layer
can be negative. It is possible that the concept of ``pressure''
in the thin layer (where the jump of $p(t,x, \varepsilon)$ took
place) one should understand in special sense. We think that the
role of $\varepsilon$ in the calculations can play so-called the
average length of free movement  of gas molecules. From our point
of view the phenomena of the formation and development of
characteristics of the  shock wave should describe in terms of
non-Archemedean distance or non-Archemedean geometry.

In addition, we can use Laguerre functions or harmonic functions
in our calculations instead of Hermite functions.

We hope that our approach will be useful for the problems of
nonlinear Optics and problems of Quantum Mechanics.

Research is partially supported by Belarussian Fundamental
Research Foundation Grant No F99M-082.

\end{document}